\documentclass[acp, manuscript]{copernicus}

\usepackage{tikz}
\usetikzlibrary{patterns}
\usetikzlibrary{shapes}
\usetikzlibrary{decorations.text}
\usetikzlibrary{shapes.multipart}
\usetikzlibrary{arrows,shapes.geometric,positioning}
\usetikzlibrary{shapes,positioning,decorations.pathmorphing}
\usetikzlibrary{calc}
\usepackage{booktabs}
\usetikzlibrary{backgrounds}
\usepackage{varwidth}
\usetikzlibrary{fit}

\usepackage{bm}
\usepackage{amsmath}
\usepackage{amsthm}
\usepackage{amssymb}
\usepackage{mathrsfs}

\usepackage{latexsym,euscript,textcomp}
\usepackage{color,graphics,epsf,dcolumn}
\usepackage{epstopdf,ulem}
\usepackage{blindtext}
\usepackage{threeparttable}
\newcommand{\beq}{\begin{equation}}
\newcommand{\eeq}{\end{equation}}

\usepackage{dcolumn}
\newcolumntype{d}[1]{D{.}{.}{#1}}

\begin{document}
\nolinenumbers

\title{Atmosphere as a steam engine}


\Author[1,2]{Anastassia M.}{Makarieva}
\Author[1,2]{Andrei V.}{Nefiodov}

\affil[1]{Theoretical Physics Division, Petersburg Nuclear Physics Institute, Gatchina  188300, St.~Petersburg, Russia}
\affil[2]{Institute for Advanced Study, Technical University of Munich, Lichtenbergstrasse 2 a, D-85748 Garching, Germany}
\affil[3]{Biotic Pump Greening Group Institute, S\~{a}o Jos\'{e} dos Campos, Brazil}


\runningtitle{Clausius-Clapeyron law and the expansion work of water vapor}

\runningauthor{Makarieva and Nefiodov}

\correspondence{A. M. Makarieva (ammakarieva@gmail.com)}

\received{}
\pubdiscuss{} 
\revised{}
\accepted{}
\published{}


\firstpage{1}

\maketitle

\begin{abstract}

Earth's atmosphere operates a steam cycle in which water vapor evaporates
from the surface, expands, condenses in colder air, and returns as
precipitation. The Clausius--Clapeyron law relates the incremental expansion
work of saturated water vapor to latent heat converted at a Carnot efficiency
corresponding to the temperature difference between evaporation and
condensation. We generalize this relation to an atmospheric column in which
condensation occurs over a range of heights and derive the expansion work per
mole of precipitated water. This includes the gravitational work associated
with lifting moist air to the mean condensation height, the expansion work
generated by condensation, and a correction for incomplete condensation.

Using GPCP v3.3 precipitation and observational constraints on condensation
height, we estimate the global steam-engine power as
$W_v=4.4\pm0.9~\mathrm{W\,m^{-2}}$. This is close to an independent
estimate of total atmospheric power,
$W=W_P+W_K\simeq4.3\pm0.6~\mathrm{W\,m^{-2}}$, obtained from the
gravitational power of precipitation and kinetic energy generation by
horizontal pressure gradients diagnosed from MERRA-2. Kinetic energy generation is
$W_K\simeq3.2\pm0.3~\mathrm{W\,m^{-2}}$, of which at least two thirds is
generated in the lower atmosphere. The smaller upper-atmospheric contribution
is dominated by temperature-related pressure gradients and is comparable to
Lorenz available potential energy generation.

We argue that the agreement between steam-engine and atmospheric power is
physically linked to condensation and precipitation fallout. By removing
water from the atmospheric gas phase and enabling column-mass redistribution,
precipitation can maintain surface pressure gradients that drive
cross-isobaric flow in the frictional lower atmosphere. The steam-engine
framework thus provides a thermodynamic basis for condensation-induced
atmospheric dynamics and identifies a major lower-atmospheric power pathway
associated with water phase transitions.

\end{abstract}

\introduction  

In a heat engine with a non-condensable working fluid, the work output equals the difference between the work done during expansion and the work required for compression. A steam engine differs in an essential way: compression of the working fluid is accomplished by condensation, so that no external mechanical work is required to return the vapor to the liquid phase. The useful work is therefore associated with the expansion of steam.

This interpretation is already contained in the Clausius--Clapeyron equation. Usually written as
\begin{equation}\label{cc}
\frac{d p_v^*}{d T} = \frac{\mathcal{L} p_v^*}{R T^2},
\end{equation}
it can also be expressed as
\begin{equation}\label{cc_reform}
\frac{R T}{p_v^*} d p_v^* = \mathcal{L} \frac{d T}{T},
\end{equation}
where $p_v^*$ is the saturated water vapor pressure, $T$ is temperature, $R$ is the universal gas constant, and $\mathcal{L}$ is the molar latent heat of vaporization (J\,mol$^{-1}$). We denote the corresponding mass-specific latent heat (J\,kg$^{-1}$) by $L=\mathcal{L}/M_v$, where $M_v$ is the molar mass of water. In this form, the equation relates the expansion work of saturated vapor to latent heat converted at Carnot efficiency.

This thermodynamic interpretation is not the usual one in modern
meteorological textbooks, where the Clausius--Clapeyron equation is
typically derived from the equality of Gibbs free energies, or chemical
potentials, of the liquid and vapor phases
\citep[e.g.,][]{iribarne1981,curry1999,andrews2010,holton2013,satoh14,bohren2023}.
The alternative Carnot-cycle derivation
\citep[e.g.,][]{brunt34,landau1967,wallace2006,feynman2011ch45}
considers an infinitesimal cycle, whose net work is
$\Delta V\,\Delta p_v^*$.
With negligible liquid volume and ideal vapor,
$\Delta V\simeq RT/p_v^*$. Equating this to $\mathcal{L}\,\Delta T/T$
gives Eq.~\eqref{cc_reform}.

Equation~\eqref{cc_reform} provides a starting point for estimating the expansion work performed by water vapor in the Earth's atmosphere. Because atmospheric condensation occurs over a range of heights and temperatures,  this work must be weighted by the amount of vapor condensing at each height. The resulting steam-engine power can then be compared with kinetic energy generation by atmospheric pressure gradients.

The atmospheric heat engine is commonly conceptualized as operating between the highest temperatures, which occur near the Earth's surface at low latitudes, and the lowest temperatures, found at high latitudes and in the upper atmosphere \citep{brunt34,pe92,bohren2023}. However, the rate of kinetic energy generation by the atmospheric heat engine remains poorly constrained. Among more recent estimates, \citet{pa12} stated that the total power associated with atmospheric motions is not known precisely but is likely on the order of $5~\mathrm{W\,m^{-2}}$. \citet{pauluis15} cited the results of \citet{laliberte15} as supporting this magnitude. In contrast, \citet{bohren2023} reported a broader range of $2$--$5~\mathrm{W\,m^{-2}}$, whereas \citet{kleidon2021} argued, on theoretical grounds, that the rate should not exceed $2~\mathrm{W\,m^{-2}}$.

A related uncertainty concerns the relative contributions of horizontal and vertical temperature differences, and their associated heat fluxes, to the generation of kinetic energy by the atmospheric heat engine. Both temperature contrasts are of order several tens of kelvin, implying a Carnot efficiency near $10\%$, a commonly cited estimate \citep{pe92,pa12,bohren2023}. The poleward heat flux, which reaches about $40~\mathrm{W\,m^{-2}}$ at its maximum, could at most sustain several watts per square metre of power. However, it is difficult to justify the existence of a Carnot cycle in this case, since the horizontal air motion connecting the effective heat source and sink is never adiabatic. In the vertical direction, the combined sensible and latent heat fluxes amount to approximately $100~\mathrm{W\,m^{-2}}$, but the efficiency of the atmospheric heat engine appears to be about a factor of two lower than the nominal Carnot value \citep{pauluis15}.

An alternative perspective was proposed already by Heinrich Hertz in 1855. He suggested that the atmosphere operates as a steam engine, with an efficiency determined by the mean temperature difference between evaporation and condensation, which he estimated to be about $15~\mathrm{K}$ \citep{mulligan1997}. For $T \simeq 300~\mathrm{K}$, this corresponds to an efficiency of approximately $5\%$. Using a latent heat flux of $80~\mathrm{W\,m^{-2}}$, corresponding to a global mean precipitation rate of $1~\mathrm{m\,yr^{-1}}$, the resulting power output of such a steam engine would be about $4~\mathrm{W\,m^{-2}}$, well within the range of reported values.

In parallel, since 2007, condensation-induced atmospheric dynamics (CIAD) has suggested that condensation and precipitation contribute substantially to atmospheric power by generating surface pressure gradients \citep[for a recent review, see][]{makarieva19}. These ideas have given rise to extensive discussion \citep{acp13,pearce20}. In particular, the relationship between condensation rate and power generation, while yielding meaningful predictions across diverse meteorological contexts, has remained largely heuristic. The steam-engine formulation developed here provides a thermodynamic framework in which this relationship can be reconsidered.

In this paper, we derive a theoretical expression for the work output of a steam engine operating in the Earth's atmosphere. We then use independent datasets---MERRA-2 for kinetic energy generation by horizontal pressure gradients \citep{gelaro2017merra2} and GPCP v3.3 for precipitation \citep{huffman2025gpcp33}---to compare the diagnosed atmospheric power with the estimated steam-engine power. We find that the two estimates are comparable in magnitude.

We then revisit previous estimates of the expansion work of water vapor, in particular those of \citet{pauluis2002b} and \citet{pa11}, and clarify how the quantities evaluated there relate to the
steam-engine work considered here. We further show that the central relationship underlying condensation-induced atmospheric dynamics expresses the correspondence between steam-engine power and kinetic energy generation in the lower atmosphere, thereby grounding CIAD in atmospheric thermodynamics.

Finally, we discuss how the diagnosed kinetic energy generation is partitioned in the vertical. More than two thirds occurs in the lower atmosphere, where mass-related pressure gradients dominate, while about one third occurs aloft, where temperature-related pressure gradients dominate. The upper-atmospheric contribution is comparable to the power associated with Lorenz available potential energy (APE), whereas the lower-atmospheric contribution is not captured by the classical Lorenz APE framework, which neglects mass-related pressure gradients by construction. We therefore argue that Lorenz APE underestimates total atmospheric power. We also discuss how condensation and precipitation fallout can help maintain surface pressure gradients and enhance kinetic energy generation in the frictional lower atmosphere. This is consistent with Hertz's observation that ``if the atmosphere were dry, the temperature differences existing in it would by themselves give rise merely to movements of minor significance'' \citep{mulligan1997}.

\section{Steam-engine efficiency and power}

\subsection{Key variables}

We use the ratio of water vapor to dry-air partial pressure, $\gamma \equiv p_v/p_d$. Air ascends from $z_1$ to $z_2$, remaining saturated throughout. Thus $p_v=p_v^*(z)$, $\gamma=\gamma(z)$, $\gamma_1\equiv\gamma(z_1)$, and $\gamma_2\equiv\gamma(z_2)<\gamma_1$. The decrease of $\gamma$ with height, $\gamma_1-\gamma_2$, measures the amount of vapor condensed during ascent. The ratio $\gamma_2/\gamma_1 \le 1$ was termed the ``incompleteness of condensation'' by \citet{jas13}. It measures the fraction of water vapor that remains uncondensed.

The expansion work per mole of condensed vapor (J\,mol$^{-1}$) is
\begin{equation}\label{Av}
A_v =
\frac{1}{\gamma_1-\gamma_2}
\int_{\gamma_2}^{\gamma_1} d\gamma
\int^{T_1}_{T(\gamma)}
\frac{dT'}{T'}\mathcal{L} 
\simeq \varepsilon_v \mathcal{L},
\end{equation}
where
\begin{equation}\label{Tc}
\varepsilon_v \equiv \frac{T_1 - T_c}{T_1}, \quad 
T_c \equiv
\frac{1}{\gamma_1-\gamma_2}
\int_{\gamma_2}^{\gamma_1} T\,d\gamma.
\end{equation}
Here $T_c$ is the mean condensation temperature. The weak temperature dependence of $\mathcal{L}$ is neglected.

Let $\mathcal{P}$ denote the molar precipitation flux (mol\,m$^{-2}\,$s$^{-1}$) and $P\equiv M_v\mathcal{P}$ the corresponding mass precipitation flux (kg\,m$^{-2}\,$s$^{-1}$). The atmospheric steam-engine power per unit surface area (W\,m$^{-2}$) is
\beq\label{Wv}
W_v = A_v \mathcal{P}
\simeq \varepsilon_v \mathcal{L}\mathcal{P}
= \varepsilon_v L P.
\eeq

Atmospheric power density is
\beq\label{W}
W = \frac{1}{\mathcal{S}}\int p \nabla \cdot \mathbf{v}\, d\Omega
= -\frac{1}{\mathcal{S}}\int \mathbf{v} \cdot \nabla p\, d\Omega,
\eeq
where $\mathbf{v}$ is air velocity, $\Omega$ is total atmospheric volume, and $\mathcal{S}$ is the planetary surface area.

Under the hydrostatic approximation, the total power can be decomposed as
\beq\label{Wsum}
W = W_P + W_K ,
\eeq
where
\beq\label{WP}
W_P \equiv M_v g (z_c-z_1)\mathcal{P}
= g(z_c-z_1)P
\eeq
is the gravitational power of precipitation, i.e., the work per unit time and surface area required to lift water from its mean evaporation height to its mean condensation height. Here $z_1$ is the mean evaporation height, and $z_c$ is the mean condensation height, taken to be equal to the mean height from which precipitation falls. The derivation of Eq.~\eqref{WP} is given in Appendix~\ref{app:WP}.

The remaining term,
\beq\label{WK}
W_K \equiv -\frac{1}{\mathcal{S}}
\int_{\Omega} \mathbf{u}\cdot\nabla p\,d\Omega ,
\eeq
where $\mathbf{u}$ is the horizontal air velocity, is the global mean kinetic energy generation by horizontal pressure gradients.

Our goal is to estimate $W = W_P + W_K$~\eqref{Wsum} and $W_v$~\eqref{Wv} from independent datasets and compare the two estimates.

\subsection{Estimating $W_v$}
\label{sec:Wv}

\begin{figure*}[tb]
\begin{minipage}[p]{0.8\textwidth}
\centerline{\includegraphics[width=1\textwidth,angle=0,clip]{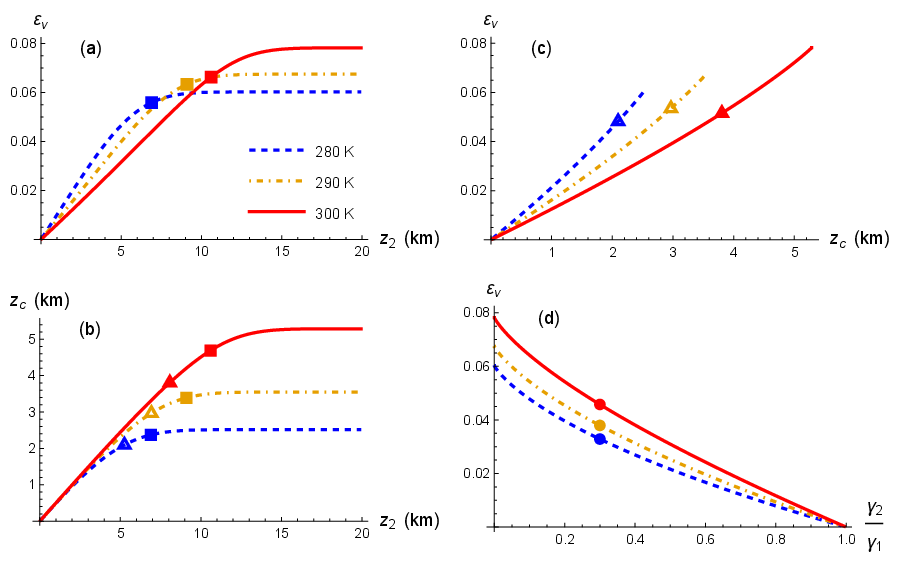}}
\end{minipage}
\caption{
Steam-engine efficiency $\varepsilon_v$ \eqref{Tc} versus condensation
top height $z_2$ (a), mean condensation height $z_c$ \eqref{zc} (c),
and incompleteness of condensation $\gamma_2/\gamma_1$ (d), calculated
for saturated ascent from the surface ($z_1=0$) with condensate removed
upon formation. Panel (b) shows $z_c$ versus $z_2$. Squares in (a) and
(b) correspond to $z_2=Z_d$ from Table~\ref{tab:Zd}. Triangles in (b) and (c) denote
$z_c=3.8$~km for $T_1=300$~K, estimated from \citet{pa12} (filled),
and $z_c$ estimated from $z_2=Z_d/1.3$ for $T_1=280$ and $290$~K
(open). Circles in (d) correspond to $\gamma_2/\gamma_1=0.3$.
}
\label{figTc}
\end{figure*}

Assuming that precipitation occurs within a moist-adiabatic column, we calculate the mean condensation temperature $T_c$ and the corresponding steam-engine efficiency $\varepsilon_v$ for different surface temperatures $T_1$ and different top heights $z_2$ of the condensation region (Fig.~\ref{figTc}a).

\begin{figure*}[tb]
\begin{minipage}[p]{0.7\textwidth}
\centerline{\includegraphics[width=1\textwidth,angle=0,clip]{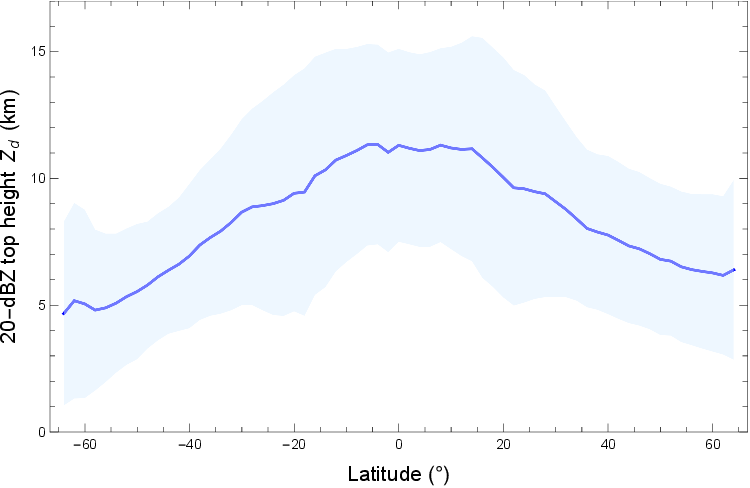}}
\end{minipage}
\caption{
Precipitation-weighted 20-dBZ top height $Z_d$ (km) versus latitude, data taken from Fig.~2 of \citet{liu2015}. Shading represents standard deviation within $2^\circ$ bins.
}
\label{fig:Zd}
\end{figure*}

We first approximate the top condensation level $z_2$ by the 20-dBZ echo-top height $Z_d$, a radar-reflectivity threshold commonly used to characterize the upper extent of the hydrometeor layer.
\citet{liu2015} reported the relative contribution to global precipitation from precipitation features with different 20-dBZ echo-top heights.
Figure~\ref{fig:Zd} shows the meridional distribution of $Z_d$ derived from their Fig.~2b. The mean precipitation-weighted $Z_d$ for the extratropics is $6.9$~km, for the tropics it is $10.6$~km, and the global mean is $9.1$~km (Table~\ref{tab:Zd}).

Assuming $T_1=300$, $280$, and $290$~K for tropical, extratropical, and
global mean precipitation, respectively, and setting $z_2=Z_d$, yields
values of $\varepsilon_v=0.056$--$0.066$, clustered around $0.06$
(Table~\ref{tab:Zd}; Fig.~\ref{figTc}a). The relatively weak dependence on $T_1$
reflects the more complete condensation at lower temperatures, which partly
compensates for the lower water vapor content. Since $Z_d$ should provide an
upper estimate for the top of the saturated region $z_2$, these values should
also represent an upper estimate of the steam-engine efficiency.

\begin{table}[t]
\centering
\caption{Precipitation fraction by region, mean 20-dBZ height $Z_d$, and estimated $\varepsilon_v$ (Fig.~\ref{figTc}) for assumed temperatures $T_1$.}
\label{tab:Zd}
\begin{tabular}{l d{1.2} c  c c c c}
\hline
Region &
\multicolumn{1}{c}{Fraction}  &
\multicolumn{1}{c}{$Z_d \pm$ s.d. (km)} &
\multicolumn{1}{c}{$T_1$ (K)} &
\multicolumn{1}{c}{$\varepsilon_v(Z_d)$} &
\multicolumn{1}{c}{$\varepsilon_v(z_c)$} &
\multicolumn{1}{c}{$\varepsilon_v(\gamma_2/\gamma_1)$} \\
\hline
Global                                               & 1.00 &  $9.1 \pm  4.3$  & 290 & 0.063 & 0.053 &  0.038\\
Tropics, $|\varphi| \leq 30^\circ$     & 0.59 & $10.6 \pm  4.2$  & 300 & 0.066 & 0.052 &  0.046\\
Extratropics, $|\varphi| > 30^\circ$ &  0.41 & $6.9 \pm  3.2$  & 280 & 0.056 & 0.048 & 0.033\\
\hline
\end{tabular}
\end{table}

For the tropical region, \citet{pa12} analyzed vertical profiles of precipitation rates and estimated the gravitational power of precipitation $W_P$~\eqref{WP} to be $1.5~\mathrm{W\,m^{-2}}$ for a latent heat flux $LP = 100~\mathrm{W\,m^{-2}}$.
This corresponds to a mean condensation height of $z_c = 3.8$~km in the tropics, which, for $T_1 = 300$~K, corresponds to $z_2 = 8.1$~km (Fig.~\ref{figTc}b).
Thus the mean tropical $Z_d=10.6$~km may overestimate the corresponding tropical $z_2=8.1$~km by a factor of about $1.3$.

If the extratropical $Z_d$ overestimates extratropical $z_2$ by the same factor, we obtain $z_2 = 5.2$~km and $z_c = 2.1$~km in the extratropics (Fig.~\ref{figTc}b). Estimated for these $z_c$, the values of $\varepsilon_v(z_c)$ cluster around $0.05$ (Table~\ref{tab:Zd} and Fig.~\ref{figTc}c).

A third, crude way of estimating $\varepsilon_v$ is to use the available information on the incompleteness of condensation, $\gamma_2/\gamma_1$ (Fig.~\ref{figTc}d), which can be related to observed precipitation efficiency. The review by \citet[][their Table~2]{sui2020} indicates that the ratio of precipitation $P$ to total condensation rate $C$ in different precipitating systems is typically $P/C \sim 0.3$--$0.5$ \citep{gamache1983,chong1989}, while the ratio of precipitation to the total flux of ascending water vapor $F$ is approximately $P/F \sim 0.2$--$0.6$. Since $C/F=(P/F)/(P/C)$, the uncondensed fraction is $1-C/F=1-(P/F)/(P/C)$. Its maximum is obtained for the smallest $P/F$ and largest $P/C$, giving $1-0.2/0.5=0.6$, while the lower bound is close to zero if most ascending vapor condenses. Thus the observations suggest a broad plausible range $0 \lesssim \gamma_2/\gamma_1 \lesssim 0.6$. We use $\gamma_2/\gamma_1=0.3$ as a representative intermediate value.

Using this value $\gamma_2/\gamma_1 = 0.3$ produces $\varepsilon_v(\gamma_2/\gamma_1)$ estimates from $0.033$ to $0.046$, depending on temperature (Fig.~\ref{figTc}d). For $T_1 = 300$~K, the obtained $\varepsilon_v(\gamma_2/\gamma_1) = 0.046$ is close to $\varepsilon_v(z_c) = 0.052$ estimated from $z_c = 3.8$~km (Fig.~\ref{figTc}c), suggesting that at least in the tropics $\gamma_2/\gamma_1 \sim 0.3$ is plausible.

The estimates summarized in Table~\ref{tab:Zd} suggest that the available evidence is consistent with a global mean steam-engine efficiency of approximately $0.04$--$0.06$.
For a global precipitation rate $P = 3.07~\mathrm{mm\,day^{-1}} = 3.55\times10^{-5}~\mathrm{kg\,m^{-2}\,s^{-1}}$ (GPCP v3.3) and $L = 2.5\times 10^6~\mathrm{J\,kg^{-1}}$, the latent heat flux is $LP = 88.8~\mathrm{W\,m^{-2}}$. Thus, with $\varepsilon_v =0.05\pm 0.01$,
\beq\label{Wvest}
W_v \equiv \varepsilon_v L P
\simeq  4.4 \pm 0.9~\mathrm{W\,m^{-2}},
\eeq
where the uncertainty reflects the range of $\varepsilon_v$ considered here.

\subsection{Estimating $W_P$}
\label{sec:WP}

Under the hydrostatic approximation, atmospheric power defined by
Eqs.~\eqref{W}--\eqref{Wsum} is partitioned into the gravitational power of
precipitation, $W_P$, and kinetic energy generation by horizontal pressure
gradients, $W_K$. We estimate these two terms separately. The term $W_P$ is
estimated directly from precipitation and condensation-height information using
Eq.~\eqref{WP}, rather than inferred from reanalysis vertical velocity.

This separation is important because estimates involving vertical velocity, or
pressure velocity $\omega \equiv Dp/Dt$, are numerically delicate. The
volume-integrated pressure-gradient work can be written in the
$\alpha\omega$ form, where $\alpha \equiv 1/\rho$ is specific volume. In this
formulation, global atmospheric power appears as a small residual of large
compensating contributions: air expands during ascent and contracts during
descent, and these contributions nearly cancel in the global mean
\citep{steinheimer2008}. For example, a vertical velocity
$w\sim 1~{\rm mm\,s^{-1}}$ gives
$|\alpha\omega|\simeq g|w|\sim 10^{-2}~{\rm W\,kg^{-1}}$. Applied to an
atmospheric column of mass $p_s/g\simeq 10^4~{\rm kg\,m^{-2}}$, this
corresponds to a column-integrated scale of order
$100~{\rm W\,m^{-2}}$, whereas the residual global atmospheric power is only a
few watts per square metre.

Since $\omega$ is itself diagnosed from mass continuity and the horizontal wind
field, small errors in the wind field, mass field, vertical interpolation, or
numerical implementation can lead to large relative errors in this residual.
As emphasized by \citet{wang2024global}, values obtained in this way are
``influenced by minor errors during computation and data quality, leading to
enormous uncertainty.'' Estimating $W_P$ directly from precipitation therefore
avoids introducing vertical-velocity uncertainty into the
precipitation-related component of atmospheric power.

For the global gravitational power of precipitation, we use $z_c=3.8$~km in
the tropics and $z_c=2.1$~km in the extratropics, with the precipitation
fractions from Table~\ref{tab:Zd}, global mean precipitation from GPCP v3.3,
and $z_1=0$. From Eq.~\eqref{WP} this gives
\beq\label{WPest}
W_P \simeq 1.1 \pm 0.3~\mathrm{W\,m^{-2}}.
\eeq
The approximate $30\%$ uncertainty reflects the broad constraint on $z_c$.
In the tropics, the adopted value $z_c=3.8$~km cannot significantly exceed
about $5$~km, which corresponds to complete condensation (Fig.~\ref{figTc}c).
A conservative lower bound is obtained by neglecting the extratropical
contribution altogether; the tropical contribution alone gives
$W_P\simeq 0.8~\mathrm{W\,m^{-2}}$. More detailed considerations of the
estimation of $W_P$ were given by \citet{jas13}.

\subsection{Estimating $W_K$}
\label{sec:WK}

The term $W_K$ is diagnosed directly from the horizontal
pressure-gradient work, Eq.~\eqref{WK}. We use MERRA-2 instantaneous
three-hourly assimilated meteorological fields on pressure levels
(\texttt{inst3\_3d\_asm\_Np}; M2I3NPASM), provided on a
$0.5^\circ \times 0.625^\circ$ latitude--longitude grid with $42$ pressure
levels and eight time slices per day. Details of the integration are provided
in Appendix~\ref{app:WK}. Among the estimates summarized in
Table~\ref{tab:WK}, this is the highest horizontal and temporal resolution
used for a global estimate of $W_K$.

For 1980--2025, the mean and standard deviation of annual values are
$W_K=3.2\pm0.07~\mathrm{W\,m^{-2}}$. Accounting for uncertainty associated
with the unresolved surface layer, we report the global estimate as
$W_K\simeq3.2\pm0.3~\mathrm{W\,m^{-2}}$.

\begin{figure*}[tb]
\begin{minipage}[p]{0.8\textwidth}
\centerline{\includegraphics[width=1\textwidth,angle=0,clip]{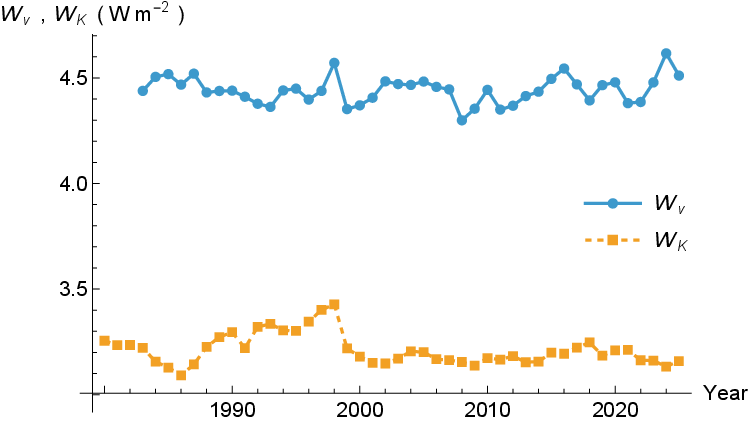}}
\end{minipage}
\caption{
Annual values of steam-engine power $W_v$, estimated from GPCP v3.3
precipitation with fixed $\varepsilon_v = 0.05$ in Eq.~\eqref{Wvest}
for $1983$--$2025$, and kinetic energy generation $W_K$, estimated from
MERRA-2 horizontal velocity and pressure-gradient fields using Eq.~\eqref{WK}
for $1980$--$2025$.
}
\label{figWvWK}
\end{figure*}

For comparison, using GPCP v3.3
precipitation and a fixed $\varepsilon_v = 0.05$ in Eq.~\eqref{Wvest} gives
$W_v = 4.4\pm 0.06~{\rm W\,m^{-2}}$ for $1983$--$2025$
(Fig.~\ref{figWvWK}). These standard deviations represent interannual
variability; the larger uncertainty in $W_v$ associated with the range of
$\varepsilon_v$ is given in Eq.~\eqref{Wvest}.

\begin{figure*}[tb]
\begin{minipage}[p]{0.8\textwidth}
\centerline{\includegraphics[width=1\textwidth,angle=0,clip]{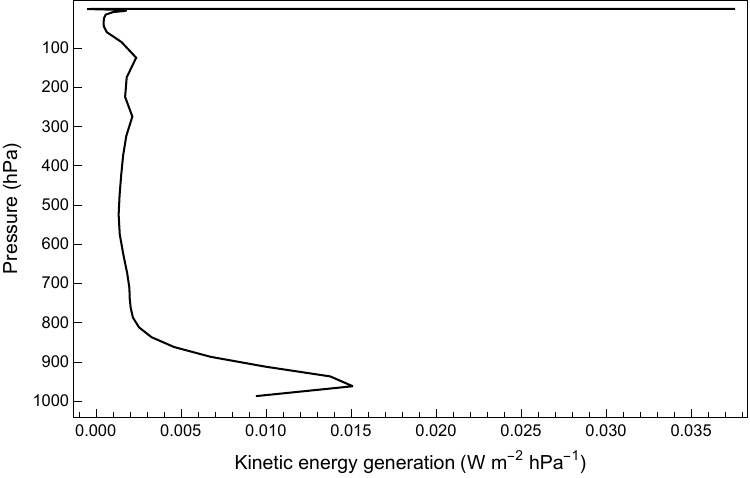}}
\end{minipage}
\caption{
Vertical profile of global kinetic energy generation for $2023$.
The resolved pressure layers contribute $3.0~\mathrm{W\,m^{-2}}$, of which
$2.2~\mathrm{W\,m^{-2}}$ is generated below $500$~hPa and
$0.8~\mathrm{W\,m^{-2}}$ above $500$~hPa. The surface-layer correction,
which adds to the lower-atmospheric contribution, depends on its treatment
(see text and Appendix~\ref{app:WK}).
}
\label{figKEvert}
\end{figure*}

The vertical distribution of $W_K$ is likewise stable throughout the analyzed
period. Figure~\ref{figKEvert} shows the profile for $2023$ as a representative
example: in individual years, kinetic energy generation consistently exhibits
a pronounced lower-atmospheric maximum, with at least two thirds of the
resolved contribution generated below $500$~hPa.

MERRA-2 provides fields on discrete pressure levels beginning at
$1000$~hPa. Consequently, the estimate of $W_K$ depends on how the unresolved
surface layer between the surface pressure $p_s$ and the lowest resolved
pressure level $p_b$ is treated. This matters because local kinetic energy
generation has a pronounced maximum in the lowest $\sim 200$~hPa
(Fig.~\ref{figKEvert}). Neglecting the surface layer reduces the global
estimate by approximately $10\%$. Assuming that
$\mathbf{u}\cdot\nabla p = 0$ at the surface and that the mean local
generation in the layer $p_s-p_b$ is half that at $p=p_b$ gives the reported
value $W_K=3.2~\mathrm{W\,m^{-2}}$. Alternatively, extrapolating
$\mathbf{u}\cdot\nabla p$ to $p_s$ from the two nearest resolved pressure
levels yields $W_K = 3.5~\mathrm{W\,m^{-2}}$ (cf.
Eqs.~\eqref{eq:surface_extrap_ck}--\eqref{eq:surface_extrap} in
Appendix~\ref{app:WK}).

\begin{table}[t]
\centering
\caption{Global kinetic energy generation $W_K$ assessed from different
reanalyses and diagnostics.}
\label{tab:WK}
\begin{tabular}{llllcl}
\toprule
Dataset & Horizontal resolution & Temporal resolution & Period & $W_K~(\mathrm{W\,m^{-2}})$ & Reference \\ 
\midrule
NCEP-R2 & $2.5^\circ\times2.5^\circ$ & daily mean & 1979--2001 & $\mathbf{\phantom{-}2.55}$ & \citet{li2007} \\
 &  & daily mean & 1979--2001 & $\phantom{-}1.74$ & \citet{marques2009} \\
 &  & inst. 6-hourly & 1979--2001 & $\phantom{-}2.58$ & \citet{marques2010} \\
 &  & inst. 6-hourly & 1979--2008 & $\mathbf{\phantom{-}1.86}$ & \citet{kim13} \\
 &  & daily mean & 1979--2013 & $\mathbf{\phantom{-}2.69}$ & \citet{pan2017earth} \\
 &  & daily mean & 1979--2014 & $-3.54$ & \citet{kim2017} \\
\addlinespace[0.35em]
ERA-40 & $2.5^\circ\times2.5^\circ$ & daily mean & 1979--2001 & $\mathbf{\phantom{-}2.06}$ & \citet{li2007} \\
 &  & daily mean & 1979--2001 & $\phantom{-}1.95$ & \citet{marques2009} \\
 &  & inst. 6-hourly & 1979--2001 & $\phantom{-}2.78$ & \citet{marques2010} \\
\addlinespace[0.35em]
JRA-25 & $2.5^\circ\times2.5^\circ$ & inst. 6-hourly & 1979--2001 & $\phantom{-}2.92$ & \citet{marques2010} \\
\addlinespace[0.35em]
MERRA & $1.25^\circ\times1.25^\circ$ & inst. 3-hourly & 1979--2008 & $\mathbf{\phantom{-}1.99}$ & \citet{kim13} \\
 &  & inst. 3-hourly & 1979--2010 & $\mathbf{\phantom{-}2.46}$ & \citet{huang15} \\
\addlinespace[0.35em]
JRA-55 & $1.25^\circ\times1.25^\circ$ & daily mean & 1979--2019 & $\phantom{-}2.34$ & \citet{Ma2021} \\
\addlinespace[0.35em]
ERA-Int & $2.5^\circ\times2.5^\circ$ & daily mean & 1979--2013 & $\mathbf{\phantom{-}2.02}$ & \citet{pan2017earth} \\
 & $1^\circ\times1^\circ$ & daily mean & 1979--2014 & $\phantom{-}3.00$ & \citet{kim2017} \\
\addlinespace[0.35em]
ERA5 & $1^\circ\times1^\circ$ & daily mean & 1985--2014 & $\phantom{-}2.93$ & \citet{wang2024global} \\
\addlinespace[0.35em]
MERRA-2 & $0.5^\circ\times0.625^\circ$ & inst. 3-hourly & 1980--2025 & $\mathbf{\phantom{-}3.20}$ & This study \\
\bottomrule
\end{tabular}
\begin{flushleft}
\footnotesize
In all studies except \citet{huang15}, $W_K$ is represented either by kinetic
energy dissipation in the Lorenz energy cycle or, where unavailable, by
conversion rates from available potential energy to kinetic energy. Boldface
values indicate estimates obtained predominantly from the horizontal
pressure-gradient work, often written in pressure coordinates as the
$\mathbf{v}\cdot\nabla Z$ form.
\end{flushleft}
\end{table}

Table~\ref{tab:WK} places our estimate in the context of previous estimates of
atmospheric kinetic energy generation. With the exception of \citet{huang15},
the quoted studies do not evaluate $-\mathbf{u}\cdot\nabla p$ directly.
Instead, they estimate the Lorenz energy cycle and report conversions from
available potential energy to kinetic energy. In the classical
Lorenz--Peixoto--Oort framework, the continuity equation contains no mass sink
associated with condensation and precipitation. The gravitational power of
precipitation is therefore absent by construction, $W_P=0$, and atmospheric
power $W$ is identified with kinetic energy generation $W_K$.

In this source-free framework, the vertical-velocity formulation
($\alpha\omega$) and the horizontal pressure-gradient formulation,
$-\mathbf{u}\cdot\nabla p$, often written in pressure coordinates as the
$\mathbf{v}\cdot\nabla Z$ form, where $Z$ is the isobaric height
(see Eqs.~\eqref{Z}--\eqref{eq:ck_local} in Appendix~\ref{app:WK}), are
formally equivalent. In a precipitating atmosphere, however, where
$W_P\ne 0$, these approaches cannot in general be expected to give identical
estimates of total atmospheric power. This point has not been considered in
discussions of discrepancies between the two formulations
\citep[e.g.,][]{marques2010,wang2024global}.

The sensitivity of vertical-motion-based estimates is apparent in
Table~\ref{tab:WK}. Some estimates based on the $\alpha\omega$ formulation
yield even negative global values \citep[e.g.,][]{kim2017}, while realistic
values have required careful tapering of boundary conditions near the surface
\citep{boer1982,marques2009}. Estimates of global atmospheric power based on
vertical velocity should therefore be treated with caution: first, because the
diagnosed power is a small residual of large terms of opposite sign
(Section~\ref{sec:WP}); and second, because the classical Lorenz framework
omits the precipitation-related term $W_P$.

The only directly comparable published estimate is that of \citet{huang15},
who diagnosed kinetic energy generation from MERRA horizontal pressure-gradient
work. Their value, $W_K\simeq2.5~\mathrm{W\,m^{-2}}$, is smaller than ours,
as expected from the coarser horizontal resolution
($1.25^\circ\times1.25^\circ$ versus $0.5^\circ\times0.625^\circ$ here)
and from the absence of an explicitly reported surface-layer correction.
Nevertheless, the qualitative vertical structure is consistent: \citet[][their
Fig.~4]{huang15} also found a pronounced maximum of kinetic energy generation
in the lower atmosphere. Thus the higher-resolution MERRA-2 estimate increases
the global magnitude of $W_K$, but preserves the central physical result that
a major part of kinetic energy generation occurs near the surface.

\begin{figure*}[tb]
\begin{minipage}[p]{0.8\textwidth}
\centerline{\includegraphics[width=1\textwidth,angle=0,clip]{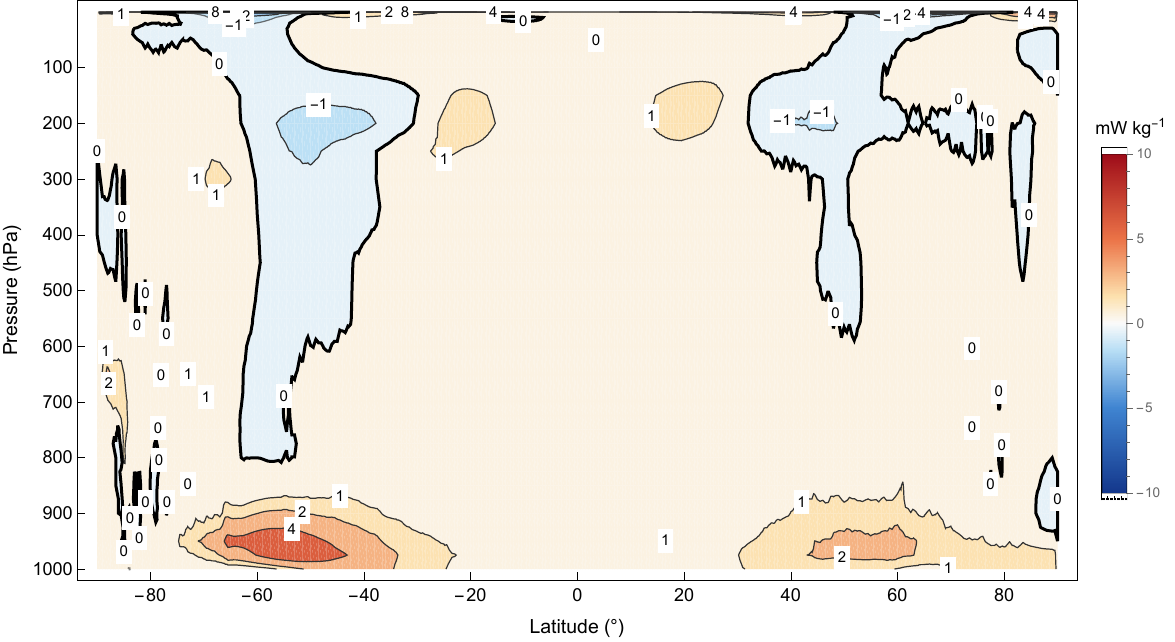}}
\end{minipage}
\caption{
Latitude--height distribution of local kinetic energy generation per unit mass,
$c_K=-\mathbf{u}\cdot\nabla p/\rho$ (Eq.~\eqref{eq:ck_local}), in
$\mathrm{mW\,kg^{-1}}$ for $2023$.
}
\label{figSpace}
\end{figure*}

Figure~\ref{figSpace} shows the characteristic latitude--height distribution
of local kinetic energy generation. In the upper troposphere, positive and
negative regions coexist. Negative values correspond to air motion toward
higher pressure and are associated with heat-pump behavior, for example in the
Ferrel cells \citep{huang14,tellus17}. In contrast, local kinetic energy generation in
the lower troposphere is positive over nearly all latitudes
(Fig.~\ref{figSpace}).

The attribution of these lower- and upper-atmospheric contributions to
different pressure-gradient components is supported by the decomposition of
\citet[][their Table~1]{tellus17}. Under hydrostatic balance, surface
pressure is proportional to the mass of the overlying atmospheric column.
For meridional circulation in both NCEP and MERRA, \citet{tellus17} found
that this surface-pressure, or mass-related, contribution dominates kinetic
energy generation in the lower atmosphere, whereas the temperature-related
contribution dominates aloft.

\section{Expansion work of atmospheric water vapor}
\label{sec:theor}

To corroborate our estimates and relate them to previous work, we now evaluate
the expansion work of atmospheric water vapor explicitly, using the left-hand
side of Eq.~\eqref{cc_reform}, $RT\,dp_v^*/p_v^*$, rather than the
temperature-side form, $\mathcal{L}\,dT/T$, used in Section~\ref{sec:Wv}.
By analogy with Eq.~\eqref{Av}, the expansion work per mole of condensed vapor
can be written as
\begin{equation}\label{INT}
A_v =
\frac{1}{\gamma_1-\gamma_2}\int_{\gamma_2}^{\gamma_1}d\gamma
\int^{z_1}_{z(\gamma)} 
\frac{dz'}{p_v^*(z')} \frac{d p_v^*}{d z'} RT(z') .
\end{equation}
Here the inner integral is taken over height, from the initial condensation
level $z_1$ to the height $z(\gamma)$ where the corresponding vapor fraction
condenses, while the outer integral averages this work over all condensing
vapor.

Since $\gamma=p_v^*/p_d$ for saturated water vapor,
\beq\label{dg}
\frac{dp_v^*}{p_v^*}
\equiv
\frac{dp_d}{p_d}+\frac{d\gamma}{\gamma}.
\eeq

Condensation enters this explicit pressure-side expression for $A_v$ in two
ways. First, in the inner integral, condensation removes water vapor from the
gas phase and thereby increases the expansion work of the remaining saturated
vapor, as Eq.~\eqref{dg} indicates. Indeed, the scale height of water vapor in
Earth's atmosphere is about $2$~km, roughly a factor of four smaller than the
atmospheric scale height \citep{weaver95}. Second, the outer integral weights
this expansion work by the amount of vapor that condenses at each height.

Assuming that all condensed moisture precipitates, $A_v$ approximately
represents the expansion work of the atmospheric steam engine per mole of
precipitated water in steady state.\footnote{This approximation arises because
the non-saturated expansion below $z_1$ is assumed to be minor and is therefore
neglected.}

Integrating Eq.~\eqref{INT} by parts gives
\begin{equation}\label{INT2}
A_v =
-\frac{1}{\gamma_1-\gamma_2}
\int_{z_1}^{z_2}
(\gamma-\gamma_2)
RT\,\frac{dp_v^*}{p_v^*}.
\end{equation}
Using Eq.~\eqref{dg}, this becomes
\begin{equation}\label{INT4}
A_v =
-\frac{1}{\gamma_1-\gamma_2}
\int_{z_1}^{z_2}
(\gamma-\gamma_2)\,RT\frac{d p_d}{p_d}
-
\frac{1}{\gamma_1-\gamma_2}
\int_{z_1}^{z_2}
\left(1-\frac{\gamma_2}{\gamma}\right)RT\,d\gamma .
\end{equation}

In hydrostatic equilibrium,
\beq\label{he}
RT\frac{dp}{p}=-Mg\,dz,
\eeq
where $M$ is the molar mass of air, assumed constant, and $g$ is the
acceleration due to gravity. Since in the vertical
$dp_d/p_d\simeq dp/p$ to accuracy $O(\gamma)$, the first term on the
right-hand side of Eq.~\eqref{INT4} becomes
\beq\label{ft}
A_g
\equiv
\frac{1}{\gamma_1-\gamma_2}
\int_{z_1}^{z_2}
(\gamma-\gamma_2)Mg\,dz
=
Mg(z_c-z_1),
\eeq
where
\beq\label{zc}
z_c
\equiv
\frac{1}{\gamma_1-\gamma_2}
\int_{\gamma_2}^{\gamma_1} z\,d\gamma
\eeq
is the mean condensation height.

Since tropospheric temperature varies by less than about $10\%$, we replace
$T$ in the second integral of Eq.~\eqref{INT4} by the condensation-weighted
mean temperature $T_c$ defined in Eq.~\eqref{Tc}. This gives
\beq\label{sec-int}
A_c
\equiv
-\frac{1}{\gamma_1-\gamma_2}
\int_{z_1}^{z_2}
\left(1-\frac{\gamma_2}{\gamma}\right)RT\,d\gamma
\simeq
RT_c
\left[
1+
\frac{\gamma_2}{\gamma_1-\gamma_2}
\ln\left(\frac{\gamma_2}{\gamma_1}\right)
\right].
\eeq
Multiplied by the partial pressure of dry air, $p_d$, the term in square
brackets corresponds to the work per unit volume performed by a hypothetical
vapor atmosphere during isothermal equilibration from the local pressure
$\gamma_2 p_d$ to the uniform pressure $\gamma_1 p_d$, cf.
Eq.~(I$a\textquotesingle$) of \citet{Margules1901}.

Combining Eqs.~\eqref{ft} and \eqref{sec-int}, we obtain
\beq\label{Av2}
\begin{aligned}
A_v &\equiv A_g + A_c
= Mg(z_c-z_1)+RT_c(1+\kappa),\\
\kappa &\equiv
\frac{\gamma_2/\gamma_1}{1-\gamma_2/\gamma_1}
\ln\left(\frac{\gamma_2}{\gamma_1}\right).
\end{aligned}
\eeq
The negative term $\kappa\le0$ accounts for the fact that vapor remaining in
the gas phase does not contribute to the useful work of the steam engine: the
expansion work associated with this noncondensed vapor must be subtracted. For
the representative value used in Section~\ref{sec:Wv},
$\gamma_2/\gamma_1=0.3$, we have $\kappa\simeq -0.52$ and hence
$1+\kappa\simeq0.48$.

In the limit $\gamma_2\to\gamma_1$, condensation vanishes. Correspondingly,
$1+\kappa\to0$, while the mean condensation height $z_c$ becomes undefined. If
no condensation occurs, water vapor behaves as an ordinary noncondensable gas.
There is then no steam engine, and separating the expansion of water vapor
from that of the other atmospheric gases has no physical meaning.

Since $1+\kappa\ge0$, Eq.~\eqref{Av2} shows that
$A_v\ge Mg(z_c-z_1)$. This lower bound exceeds the gravitational work required
to lift one mole of water, $M_vg(z_c-z_1)$, by the factor
$M/M_v\simeq1.6$. Thus the steam-engine power $W_v$, Eq.~\eqref{Wv}, is always
larger than the gravitational power of precipitation $W_P$, Eq.~\eqref{WP}, by
at least this factor.

For complete condensation, $\gamma_2/\gamma_1\to0$ and hence $\kappa\to0$. If,
in addition, the mean evaporation height is small compared with the mean
condensation height, $z_1\ll z_c$, the steam-engine power becomes
\beq\label{Wv2}
W_v=A_v\mathcal{P}=(Mgz_c+RT_c)\mathcal{P}.
\eeq

\begin{figure*}[tb]
\begin{minipage}[p]{0.8\textwidth}
\centerline{\includegraphics[width=1\textwidth,angle=0,clip]{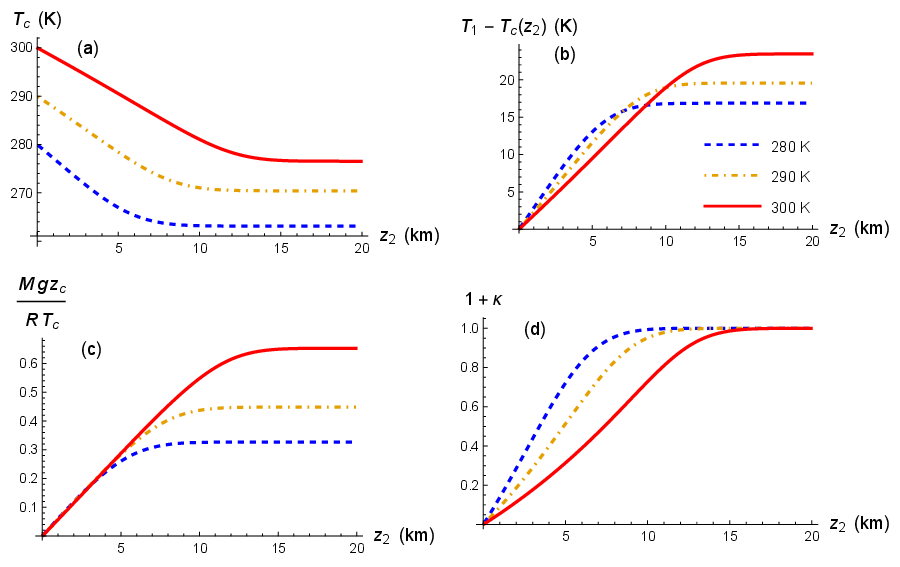}}
\end{minipage}
\caption{
Parameters of Eq.~\eqref{Av2} calculated for saturated ascent from the
surface ($z_1=0$) with condensate removed upon formation, for
$T_1=280$, $290$, and $300$~K.
}
\label{figAv}
\end{figure*}

\section{Previous work}

\subsection{\citet{pauluis2002b}}
\label{sec:Wvap}

\citet{pauluis2002b} emphasized the essential feature of the steam engine:
``In the classic Carnot cycle, work is produced by a combination of a warm expansion and cold compression. In contrast, the work performed by water vapor results primarily from its expansion during its ascent: the condensed water, being incompressible, does not require any compression work to be brought to the surface. This explains that although water vapor accounts for less than 2\% of the total mass of the atmosphere, it may produce a large fraction of the total work.''

They defined $W_{\rm vap}$ as the ``total amount of work produced by water vapor''. Written per unit surface area, this quantity is
\beq\label{Wvap}
W_{\rm vap}
=
\frac{1}{\mathcal{S}}\int p_v \nabla \cdot \mathbf{v}\, d\Omega .
\eeq
Neglecting boundary fluxes, Eq.~\eqref{Wvap} can be written equivalently as
\beq\label{Wvapgrad}
W_{\rm vap}
=
-\frac{1}{\mathcal{S}}\int \mathbf{v}\cdot\nabla p_v \, d\Omega .
\eeq

Although $W_{\rm vap}$ was defined as total work, \citet{pauluis2002b} subsequently referred to it as the expansion work of water vapor. However, because the integration is over the entire atmosphere, $W_{\rm vap}$ also includes a term of opposite physical character: the increase of water vapor partial pressure $p_v$ during evaporation at the Earth's surface. This term can be interpreted formally as a ``contraction work'' of water vapor. Accordingly, the expression obtained by \citet{pauluis2002b} contains an additional negative contribution compared with Eq.~\eqref{Wv2} (see Appendix~\ref{app:Wvap} for the derivation):
\beq\label{WvapP}
W_{\rm vap}
=
\left[Mgz_c+R(T_c-T_b)\right]\mathcal{P},
\eeq
where $\mathcal{P}$ is the molar precipitation flux, $T_b$ is the mean temperature of evaporation, the ``boiler'', and $T_c$ is the mean temperature of condensation, the ``condenser''.

Since $T_b>T_c$, with $0<T_b-T_c\ll T_c$, the temperature contribution
in Eq.~\eqref{WvapP} is negative and relatively small. For representative
values $T_c \simeq 270$~K and $T_b-T_c \simeq 17$--$20$~K
(Fig.~\ref{figAv}a,b), this contribution is about $-0.07\,RT_c$,
whereas $Mgz_c \simeq 0.45\,RT_c$ (Fig.~\ref{figAv}c). Thus, per mole
of precipitated water, $A_{\rm vap}\equiv W_{\rm vap}/\mathcal{P}$ is
about one quarter to one third of
$A_v=Mgz_c+RT_c$, the expansion work of saturated water vapor.
Accordingly, \citet{pauluis2002b} concluded that $W_{\rm vap}$ should
amount to only ``20\%--30\% of the work of a perfect heat engine that
would transport the latent heat.''

Thus $A_{\rm vap}$ remains comparable to the gravitational work required to lift precipitating water. This result has supported the view that most of the work potentially available from the vertical latent-heat flux is consumed locally by moist convection and the hydrological cycle, leaving little to generate large-scale atmospheric motion \citep[e.g.,][]{kleidon2021}.

This conclusion, however, follows from a definition of $W_{\rm vap}$ that
includes a substantial negative contribution associated with the increase
of $p_v$ during evaporation. Physically, this term represents work supplied
by the liquid-to-vapor phase transition, as water enters the gaseous phase
with molar volume $RT/p_v$, rather than work generated by the vapor itself.
The increase of $p_v$, which formally
plays the role of compression, results not from compression of the same
gaseous working substance, but from its replenishment from the liquid
reservoir. Thus Eq.~\eqref{Wvap} combines the positive expansion work generated during
condensation with an opposing contribution associated with replenishment of
vapor by evaporation. It therefore represents a different quantity from the
condensation-region expansion work $A_v$ considered here. Starting from the
Clausius--Clapeyron relation in its Carnot-cycle form,
Eq.~\eqref{cc_reform}, isolates the latter quantity directly.

\subsection{Condensation-induced atmospheric dynamics}
\label{sec:ciad}

The condensation-induced atmospheric dynamics (CIAD) approach originated from
the idea that the non-equilibrium vertical distribution of water vapor partial
pressure, maintained by condensation, can generate atmospheric power
proportional to the condensation rate \citep[for an overview, see][]{makarieva19}.
The proposed local power density was
\beq\label{Wloc}
\sigma_{\rm CIAD} \equiv
-w \left(
\frac{\partial p_v}{\partial z}
-
\frac{p_v}{p}\frac{\partial p}{\partial z}
\right)
=
-wp\frac{\partial \gamma_p}{\partial z},
\eeq
where $w$ is vertical velocity and $\gamma_p\equiv p_v/p$.

To connect this expression with the steam-engine formulation, we express the
condensation rate explicitly. Let
$\gamma\equiv p_v/p_d=\mathcal{N}_v/\mathcal{N}_d$, where $p_d$ is dry-air
partial pressure and $\mathcal{N}_v$ and $\mathcal{N}_d$ are the molar
densities of water vapor and dry air. In a condensation region, where dry air
is conserved and water vapor is removed from the gas phase, the continuity
equations give
\beq\label{Nc}
\dot{\mathcal{N}}_c
=
-\mathcal{N}_d\frac{D\gamma}{Dt},
\eeq
where $\dot{\mathcal{N}}_c$ is the molar condensation rate per unit volume,
defined positive when vapor is removed from the gas phase. Since condensation
in precipitating systems is primarily associated with vertical motion of moist
air, we write, to leading order,
\beq\label{Ncvert}
\dot{\mathcal{N}}_c
\simeq
-\mathcal{N}_d w\frac{\partial\gamma}{\partial z}.
\eeq
Multiplying Eq.~\eqref{Ncvert} by $RT$ and using
$\mathcal{N}_dRT=p_d$, we obtain
\beq\label{RTNc}
RT\dot{\mathcal{N}}_c
\simeq
-wp_d\frac{\partial\gamma}{\partial z}.
\eeq
Thus, up to this leading-order vertical-motion approximation and the
replacement of $p_d$ by $p$ and $\gamma$ by $\gamma_p$ in the dilute-vapor
limit, the local CIAD expression corresponds approximately to the local power
density $RT\dot{\mathcal{N}}_c$.

Integrating this local power density over the condensation layer gives the
condensational power per unit horizontal area,
\beq\label{WcCIAD}
W_c
=
\int_{z_1}^{z_2}RT\dot{\mathcal{N}}_c\,dz.
\eeq
Replacing $T$ by the condensation-weighted mean temperature $T_c$ gives
\beq\label{WcCIAD2}
W_c
\simeq
RT_c\int_{z_1}^{z_2}\dot{\mathcal{N}}_c\,dz
=
RT_c\mathcal{P},
\eeq
where $\mathcal{P}$ is the molar precipitation flux, assuming that condensed
water precipitates. In the complete-condensation limit,
$\gamma_2/\gamma_1\to 0$ and hence $\kappa\to0$, so that
$A_c=RT_c$. Equation~\eqref{WcCIAD2} is then precisely the condensational
contribution $A_c\mathcal{P}$ in Eq.~\eqref{Av2}.

The subtraction term in Eq.~\eqref{Wloc} clarifies why the CIAD expression
isolates this condensational contribution. It removes the hydrostatic
contribution associated with the vertical decrease of air pressure. If water
vapor behaved as a passive noncondensable constituent with vertically constant
$\gamma_p$, the bracket in Eq.~\eqref{Wloc} would vanish. The nonzero CIAD
term therefore measures the departure from passive hydrostatic behavior caused
by condensation.

Thus the local CIAD power density corresponds to the complete-condensation
limit of the condensational expansion-work term $A_c\mathcal{P}$, with the
hydrostatic/gravitational contribution subtracted. The full steam-engine
formulation restores the total expansion work by adding the gravitational
contribution $A_g\mathcal{P}=Mg(z_c-z_1)\mathcal{P}$ and by allowing for
incomplete condensation through $A_c=RT_c(1+\kappa)$. In this sense, the
steam-engine formulation provides a thermodynamic grounding for the earlier
CIAD relation between condensation rate and atmospheric power generation.

If the local power $\sigma_{\rm CIAD}$ is identified, in order of magnitude,
with local kinetic energy generation by horizontal pressure gradients,
$\sigma_{\rm CIAD}\sim -\mathbf{u}\cdot\nabla p$, dimensional analysis gives a
characteristic horizontal pressure difference. Taking $u/w\sim L/H$,
$|\nabla p|\sim \Delta p/L$, and
$|\partial\gamma/\partial z|\sim \Delta\gamma/H$, one obtains
$\Delta p\sim p\Delta\gamma$. For nearly complete condensation,
$\Delta\gamma\sim\gamma_1$, and hence $\Delta p$ is of the order of the
near-surface water vapor partial pressure, $p\gamma_1\sim 10$~hPa.

\citet{hess07} noted that this estimate is close to the pressure differences
observed in large-scale circulations with $L\sim10^3$~km. For example, in the
Hadley circulation, the horizontal surface pressure difference is about
$12$~hPa \citep{tellus17}. \citet{holton04} similarly noted that horizontal
pressure differences of order $10$~hPa occur in systems of widely different
spatial scales, including tornadoes, squall lines, and hurricanes. 

Within the CIAD framework, this relative constancy is interpreted as a
manifestation of the constraint that condensation imposes on kinetic energy
generation. For tropical cyclones, equating the condensation-domain integral
of $\sigma_{\rm CIAD}$ to boundary-layer kinetic energy generation, and
applying cyclostrophic balance, yields an upper limit on maximum wind
velocity \citep{mn25}.

The relation between the steam-engine formulation and CIAD can now be stated
more specifically. In the complete-condensation limit, the earlier CIAD
expression isolates the condensational contribution to steam-engine power,
$A_c\mathcal{P}=RT_c\mathcal{P}$, and associates it with kinetic energy
generation by horizontal pressure gradients. This correspondence is
consistent with the diagnosed vertical distribution of $W_K$
(Section~\ref{sec:WK}): a major part of kinetic energy generation occurs in
the lower atmosphere, where the surface-pressure, or column-mass-related,
contribution dominates.

\begin{figure}[t]
\centering
\begin{tikzpicture}[
    x=1.35cm,
    y=1cm,
    font=\small,
    barlabel/.style={font=\small},
    lefttext/.style={font=\small, align=right},
    symbollabel/.style={font=\large}
]

\def\WP{1.1}
\def\WK{3.2}
\def\Wtot{4.3}
\def\GL{1.2}
\def\Wg{1.8}
\def\Wc{2.6}
\def\Wv{4.4}

\def\WKupper{1.2}
\def\WKlower{2.0}

\def\h{0.56}
\def\yW{2.4}
\def\yL{1.2}
\def\yV{0.0}

\definecolor{gravPower}{RGB}{92,157,214}
\definecolor{kinPower}{RGB}{237,125,49}
\definecolor{lorenzPower}{RGB}{112,173,71}
\definecolor{wgPower}{RGB}{165,105,189}
\definecolor{wcPower}{RGB}{244,180,0}

\node[lefttext, anchor=east] at (-1.15,\yW+0.5*\h) {Atmospheric power};
\node[lefttext, anchor=east] at (-1.15,\yL+0.5*\h) {Lorenz APE generation};
\node[lefttext, anchor=east] at (-1.15,\yV+0.5*\h) {Steam-engine power};

\node[symbollabel, anchor=east] at (-0.18,\yW+0.5*\h) {$W$};
\node[symbollabel, anchor=east] at (-0.18,\yL+0.5*\h) {$G_L$};
\node[symbollabel, anchor=east] at (-0.18,\yV+0.5*\h) {$W_v$};

\fill[gravPower] (0,\yW) rectangle (\WP,\yW+\h);
\fill[kinPower!45] (\WP,\yW) rectangle (\WP+\WKupper,\yW+\h);
\fill[kinPower!85!black] (\WP+\WKupper,\yW) rectangle (\Wtot,\yW+\h);

\node[barlabel, above] at (0.5*\WP,\yW+\h) {$W_P=1.1 \pm 0.3$};
\node[barlabel, above] at (\WP+0.5*\WK,\yW+\h) {$W_K=3.2\pm 0.3$};

\node[barlabel] at (\WP+0.5*\WKupper,\yW+0.5*\h) {upper};
\node[barlabel, text=white] at (\WP+\WKupper+0.5*\WKlower,\yW+0.5*\h) {lower};

\node[barlabel, right] at (\Wtot,\yW+0.5*\h) {$4.3\pm 0.6$};

\fill[lorenzPower] (0,\yL) rectangle (\GL,\yL+\h);
\node[barlabel, right] at (\GL,\yL+0.5*\h) {$1.2$};

\fill[wgPower] (0,\yV) rectangle (\Wg,\yV+\h);
\fill[wcPower] (\Wg,\yV) rectangle (\Wv,\yV+\h);

\node[barlabel, below] at (0.5*\Wg,\yV) {$W_g=1.6 W_P$};
\node[barlabel, below] at (\Wg+0.5*\Wc,\yV) {$W_c=2.6$};

\node[barlabel, right] at (\Wv,\yV+0.5*\h) {$4.4 \pm 0.9$};

\end{tikzpicture}

\caption{Comparison of global atmospheric power estimates
($\mathrm{W\,m^{-2}}$). Atmospheric power is partitioned as
$W=W_P+W_K$, and steam-engine power as $W_v=W_g+W_c$. The uncertainty
of $W$ is conservatively estimated as the sum of the uncertainties in
$W_P$ and $W_K$. The partition of $W_K$ indicates that at least two thirds is generated below
$500$~hPa; its precise lower-atmospheric value depends on the treatment of
the unresolved surface layer. The value of $G_L$ is the observational estimate of
\citet{romanski2013}; see Section~\ref{sec:lorenz}. }
\label{fig:budget}
\end{figure}
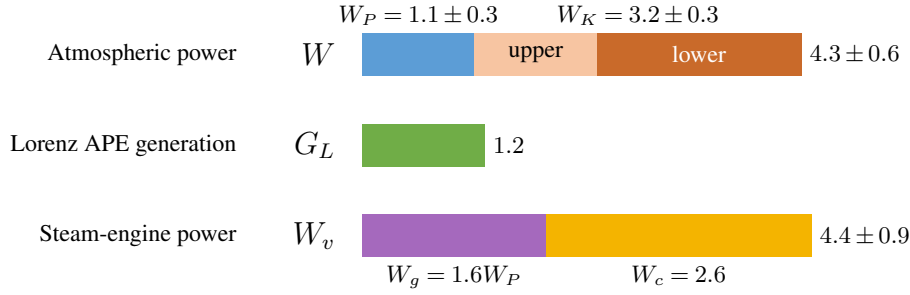

Two comparisons are therefore relevant. The full steam-engine power $W_v$ is
close to total atmospheric power $W=W_P+W_K$ (Fig.~\ref{fig:budget}). More specifically, its
condensational component $A_c\mathcal{P}$ is comparable to the
lower-atmospheric contribution to $W_K$, which constitutes a major part of
total kinetic energy generation and is dominated by mass-related pressure
gradients. This specific correspondence indicates that the agreement between
the total powers is not merely numerical, but is associated with a physically
identifiable lower-atmospheric mechanism. The implications of this result for
the Lorenz APE framework are considered in Section~\ref{sec:lorenz}.

\subsection{\citet{pa11}}
\label{sec:pa11}

Whereas \citet{pauluis2002b} explicitly sought to quantify the expansion work
performed by water vapor, \citet{pa11} considered the net mechanical work and
efficiency of prescribed moist thermodynamic cycles. The two quantities need
not coincide. The expansion work generated in saturated condensing motion is
a positive contribution, while the net work of a complete cycle includes all
branches required to close the circulation and may therefore contain
compensating negative contributions.

\citet{pa11} showed that the efficiency of an idealized steam cycle decreases
relative to the Carnot efficiency as relative humidity decreases. For
Earth-like conditions, the efficiency of the totally unsaturated steam cycle
was estimated to be only about one-sixth to one-fifth of the Carnot value.
This result is not in conflict with substantial condensation-related kinetic
energy generation in the lower atmosphere proposed within the CIAD framework.
A moist circulation may generate considerable positive work in its lower
branch while having a low net efficiency because this work is partly consumed
in closing the circulation aloft.

To illustrate this point, consider a circulation with horizontally isothermal
lower and upper branches, in which condensation maintains a surface pressure
difference of magnitude
\beq
\frac{\Delta p_s}{p_s}=\gamma ,
\eeq
where $\gamma$ is the near-surface molar fraction of water vapor. To first
order in $\gamma$, flow toward lower pressure along the warm lower branch at
temperature $T_b$ generates positive kinetic energy $RT_b$ per mole of
precipitated water, comparable to the condensational expansion work $A_c$.
If the circulation is closed by return flow along a colder upper isothermal
branch at temperature $T_c$, the latter proceeds toward higher pressure and
performs negative work $-RT_c$. The net kinetic energy generation of the
cycle, per mole of precipitated water, is then, taking
$T_b=290~\mathrm{K}$,
\beq
A_{K}=R(T_b-T_c),
\quad
\varepsilon_{K}
=
\frac{R(T_b-T_c)}{\mathcal{L}}
=
\frac{RT_b}{\mathcal{L}}\varepsilon_C
\simeq 0.05\,\varepsilon_C
\ll \varepsilon_C .
\eeq

This value refers to kinetic energy generation only. In the presence of phase
transitions, total atmospheric power also includes the gravitational power of
precipitation, $W_P$ \citep[see Eq.~(A3) of][]{tellus17}. With
$W_P\simeq1.1~\mathrm{W\,m^{-2}}$ and a latent heat flux
$Q_L\simeq90~\mathrm{W\,m^{-2}}$, the gravitational contribution corresponds
to
\beq
\frac{W_P}{Q_L}\simeq0.012
\simeq0.12\,\varepsilon_C ,
\eeq
for a characteristic atmospheric Carnot efficiency
$\varepsilon_C\simeq0.1$. Thus the steam-cycle efficiency in this limiting
case is approximately
\beq
\varepsilon_P
\simeq
(0.05+0.12)\varepsilon_C
\simeq0.17\,\varepsilon_C ,
\eeq
which is within the one-sixth to one-fifth Carnot range estimated by
\citet{pa11} for Earth-like conditions.\footnote{Isothermal expansion along
the warm lower branch requires an additional heat input of order
$\gamma RT_b$. Relative to the latent heat input $\gamma\mathcal{L}$, this
correction is $RT_b/\mathcal{L}\simeq0.05$ and does not affect the present
order-of-magnitude comparison.}

Thus low net cycle efficiency can coexist with significant positive kinetic
energy generation in the condensation-driven lower branch: the low value
reflects compensation by negative kinetic energy generation aloft.

Such compensation is dynamically possible where air possesses sufficient
kinetic energy to move against an adverse pressure gradient. Neglecting other
forces, crossing a pressure rise $\Delta p$ consumes kinetic energy of order
$\Delta p/\rho$ and therefore requires
$\rho v^2/2\gtrsim\Delta p$. Figure~\ref{figSpace} shows regions of negative
kinetic energy generation in the upper atmosphere, where high-velocity air
moves toward higher pressure and loses part of its kinetic energy. Tropical
cyclones provide a concrete analogue: \citet{kurihara75} and
\citet{smith18} found strong kinetic energy generation in low-level inflow
together with negative kinetic energy generation in upper-level outflow,
where rapidly moving air proceeds outward against the radial pressure-gradient
force.

The horizontally isothermal cycle is therefore a useful limiting case. It is
not representative, however, of a circulation in which kinetic energy
generated in the frictional lower atmosphere is largely dissipated locally,
such that $\rho v^2/2\ll\Delta p_s$. This condition is characteristic of
Hadley-type circulations, where surface pressure differences can be of order
$10$~hPa, while the kinetic energy of the ascending air is insufficient to
sustain upper-level outflow against a comparable adverse pressure difference
\citep[cf.][]{tellus17}. In such a circulation, the pressure difference must
become small near the upper branch.

Let $z_e$ denote the isobaric height, defined by $\Delta p(z_e)=0$. The
relation between surface pressure and temperature differences depends on
$z_e$ \citep{jcli15}. Writing this relation, in magnitude, as
\beq
\frac{\Delta T}{T}
\simeq
k\,\frac{\Delta p_s}{p_s},
\eeq
the coefficient $k$ is, in the simplest hydrostatic estimate, given by
$k=H/z_e$, where $H$ is the atmospheric scale height. Thus $k=1$ for an
isobaric height equal to one scale height, while an isobaric level near
$500$~hPa, dividing the atmospheric mass approximately in half, corresponds
to $k\simeq1/\ln 2\simeq1.4$.

Since the condensation-maintained surface pressure difference is of order
$\Delta p_s/p_s\sim\gamma$, it follows that
\beq
\frac{\Delta T}{T}\sim k\gamma .
\eeq
The sensible and latent heat inputs associated with this circulation then
scale as
\beq
Q_s\sim C_p\Delta T\sim k\gamma C_pT,
\qquad
Q_L\sim\gamma\mathcal{L},
\eeq
and hence
\beq
B\equiv\frac{Q_s}{Q_L}
\sim
k\,\frac{C_pT}{\mathcal{L}}
\simeq0.2k .
\eeq
Taking $k\simeq1.4$ gives, approximately, $B\simeq0.3$.

This scaling connects directly to the mixed steam--Carnot cycle considered by
\citet{pa11}. Denoting the work output of this mixed cycle by $W_{\rm mix}$,
the relation given by \citet{pa11} can be written as
\beq
W_{\rm mix}=\varepsilon_P Q_L+\varepsilon_C Q_s ,
\qquad
\frac{\varepsilon_{\rm mix}}{\varepsilon_C}
=
\frac{\varepsilon_P/\varepsilon_C+B}{1+B}.
\eeq
Taking $\varepsilon_P/\varepsilon_C\simeq0.2$ and $B\simeq0.3$ gives
\beq
\frac{\varepsilon_{\rm mix}}{\varepsilon_C}
\simeq0.4 .
\eeq
For a total heat input $Q_L+Q_s\sim10^2~\mathrm{W\,m^{-2}}$ and
$\varepsilon_C\sim0.1$, this corresponds to
$W_{\rm mix}\sim4~\mathrm{W\,m^{-2}}$, close to the atmospheric
power diagnosed here. Given the approximate scalings involved, this agreement
should be interpreted as consistency in magnitude rather than a precise
quantitative prediction.

Importantly, the temperature contrast entering this estimate is not an
independent addition to the steam-engine cycle. It follows from the dynamical
constraint imposed by the steam-engine mechanism: if condensation-maintained
pressure gradients generate kinetic energy in the frictional lower
atmosphere, while extensive cancellation of this work by motion against an
adverse pressure gradient aloft is avoided, the upper-level pressure
difference must become small or reverse sign. In a hydrostatic atmosphere,
this requires a temperature contrast. Thus an efficiency of terrestrial
magnitude follows naturally once the dynamical consequences of the
steam-engine mechanism are included.

In this sense, the mixed-cycle formulation of \citet{pa11} and the present
steam-engine analysis are complementary. The former describes how latent and
sensible heat inputs determine the net work of a moist cycle; the latter
identifies why both inputs arise in the terrestrial circulation.

\section{Link to Lorenz available potential energy}
\label{sec:lorenz}

We now compare the steam-engine formulation with Lorenz available potential
energy (APE), denoted here by $A_L$. In its most general form, abstracting from
the details of any particular formulation, the Lorenz APE framework can be
summarized as follows.

First, $A_L$ is defined as a function of the atmospheric state. Second, energy
conservation and the continuity equation are used to derive an evolution
equation \citep[see, e.g., Eq.~(16) of][]{lorenz55}:
\begin{equation}
\label{AL}
\frac{\partial A_L}{\partial t} = G_L - W_K ,
\end{equation}
where $W_K$ is kinetic energy generation by pressure gradients
\citep[cf. Eq.~(17) of][]{lorenz55}, and $G_L$ is APE generation by atmospheric
processes. Since $G_L$ is expressed in terms of the atmospheric state and the
corresponding thermodynamic tendencies, it can, in principle, be diagnosed
independently of the kinetic-energy budget \citep[cf. Eq.~(18) of][]{lorenz55}.
In a statistically steady state, $\partial A_L/\partial t=0$, so that
$G_L=W_K$.

Much discussion of Lorenz APE concerns the definition of $A_L$ itself. For
example, an important question is whether APE should be defined relative to a
static reference state or relative to a dynamically constrained atmospheric
state \citep[e.g.,][]{Shepherd1993,Marquet1995}. However, if an APE formulation is intended to represent the full potential energy source of atmospheric kinetic energy, its generation term $G_L$ must
be consistent with independently diagnosed global kinetic energy generation
$W_K$ in steady state. Since interannual variability in $W_K$ is small compared with the uncertainty
in its absolute value (Fig.~\ref{figWvWK}), the practical challenge is to
obtain an independent diagnostic of the climatological mean value of $W_K$.

Lorenz APE was defined as the difference between the potential energy of the
actual atmosphere and that of a stably stratified reference state obtained by
adiabatic rearrangement and free of horizontal pressure gradients. Its
interpretation was therefore linked from the outset to horizontal pressure
forces arising from nonuniform stratification. As \citet{lorenz55} noted for
spatially nonuniform cooling, ``The cooling removes total potential energy
from the system, but it still disturbs the stratification, thus creating
horizontal pressure forces which may convert total potential energy into
kinetic energy.''

Horizontal pressure gradients, however, may also arise without horizontal
temperature contrasts: in an isothermal atmosphere, horizontal variations in
column mass are sufficient. Such gradients are not represented by Lorenz APE,
which is zero in this case because isentropic and isobaric surfaces coincide,
eliminating the pressure variance on isentropic surfaces entering its
definition \citep[see Eq.~(4) of][]{lorenz55}. Consistently, \citet{lorenz1960} stated that without horizontal temperature
contrast there would be no generation of available potential energy.

In addition to excluding pressure gradients associated with horizontal
variations in column mass rather than temperature, the derivation of
Eq.~\eqref{AL} uses a continuity equation without sources or sinks of
atmospheric mass.\footnote{In this source-free case, the water-lifting power
$W_P$ is zero, and kinetic energy generation $W_K$ equals total atmospheric
power $W$, Eq.~\eqref{W}.} To illustrate why this matters, consider an
isothermal atmosphere, for which
\[
\frac{1}{\rho}\nabla \rho = \frac{1}{p}\nabla p .
\]
In two dimensions, source-free continuity gives
\begin{equation}
-u\frac{\partial p}{\partial x}
=
p\left(
\frac{\partial u}{\partial x}
+
\frac{\partial w}{\partial z}
\right)
+
w\frac{\partial p}{\partial z}.
\label{eq:upx_continuity}
\end{equation}
If the atmosphere is hydrostatic and isothermal, then
\[
\frac{\partial p}{\partial z}=-\frac{p}{H},
\]
where $H$ is the pressure scale height, and Eq.~\eqref{eq:upx_continuity}
becomes
\begin{equation}
-u\frac{\partial p}{\partial x}
=
p\left(
\frac{\partial u}{\partial x}
+
\frac{\partial w}{\partial z}
-
\frac{w}{H}
\right).
\label{eq:upx_scale_form}
\end{equation}
Dividing by $p$ gives the scale estimate
\begin{equation}
-\frac{u_*}{L}\frac{\Delta p}{p}
\sim
\frac{u_*}{L}
+
\frac{w_*}{h}
-
\frac{w_*}{H},
\label{eq:upx_scale}
\end{equation}
where $u_*$ and $w_*$ are characteristic horizontal and vertical velocities,
$L$ and $h$ are the horizontal and vertical scales of motion, and
$\Delta p/p$ is the characteristic relative horizontal pressure perturbation.

In the lower atmosphere, $\Delta p/p \sim \gamma \sim 10^{-2}$. Thus the term
directly associated with kinetic energy generation appears in the continuity
equation as a small residual of larger terms. If the physical mechanism
responsible for kinetic energy generation is absent from the governing
equations, the continuity equation may still be satisfied with high precision
while the diagnosed kinetic energy generation is incorrect. Conversely, once
the continuity equation contains source--sink terms, the kinetic-energy budget
can acquire pressure-work terms that are not captured by Lorenz APE.

Lorenz's own quantitative estimate points in the same direction. When
\citet{lorenz1960} used available estimates of differential heating to
diagnose $G_L$, he obtained an APE generation rate about a factor of two
smaller than independent estimates of atmospheric kinetic energy generation
from velocity and pressure fields. He therefore examined additional
correlations in the heating distribution, including latitudinal variations in
albedo, which increase absorbed solar radiation in warmer low latitudes
relative to colder high latitudes.

A closely related result was later obtained by \citet{romanski2013}. Using
mostly satellite-derived estimates of diabatic heating components, including
latent heating, radiative flux convergence, and surface sensible heat flux,
they diagnosed a global Lorenz APE generation $G_L$ of about
$1.2~{\rm W\,m^{-2}}$. This value is substantially smaller than independent
estimates of global kinetic energy generation $W_K$, which are typically of
order $2$--$3~{\rm W\,m^{-2}}$ at comparable resolution
(Table~\ref{tab:WK}).

Although \citet{romanski2013} did not relate their estimate to the discrepancy
identified by \citet{lorenz1960}, their result reproduces it in modern
observational form: $G_L$, diagnosed from differential heating alone, falls
short of independently diagnosed kinetic energy generation $W_K$, and still
further short of total atmospheric power $W$, which in a precipitating
atmosphere also includes the water-lifting term $W_P$ (Fig.~\ref{fig:budget}).
To our knowledge, no later observational recalculation has superseded their
estimate.\footnote{A higher estimate was obtained by \citet{ahbe2017} using
the NCAR CESM1.0.4 climate model, although they did not cite the observational
analysis of \citet{romanski2013}. This estimate should be interpreted as
model-based: the same model atmosphere has a comparatively large kinetic
energy generation rate, about $4~{\rm W\,m^{-2}}$, exceeding typical
observational or reanalysis-based estimates at similar spatial resolution.}
Most assessments of the Lorenz energy cycle, including those listed in
Table~\ref{tab:WK}, assume that Eq.~\eqref{AL} is an identity and estimate
$G_L$ from $W_K$ rather than independently. Generation terms expressed through
differential heating are generally recognized to be highly sensitive to
reanalysis configuration \citep[][]{kim2017}.

The discrepancy $G_L<W$ suggests that not all kinetic energy generation needs
to originate from differential heating as represented in the classical Lorenz
framework. It points to the need to examine physically distinct pathways for
generating kinetic energy. One such pathway is the steam-engine mechanism
considered here.

\section{Potential energy transformation in the presence of a steam engine}
\label{sec:mech}

We now consider how the expansion work of water vapor in the Earth's
atmosphere can be converted into kinetic energy generation in the lower
atmosphere. Modern atmospheric models clarify the sequence of processes involved in this
mechanism by making explicit the coexistence of slow resolved motion, rapid pressure adjustment,
and moist physical processes.

A concrete example is the Advanced Research core of the Weather Research and
Forecasting model, WRF-ARW \citep{skamarock2019wrf}. In WRF-ARW, the
resolved dynamical equations are advanced over a finite time step, while fast
acoustic modes are treated on shorter substeps. Moist physical processes,
including cloud microphysics and associated phase transitions, are applied
separately from the main dynamical update.

Microphysics is applied at the end of the time step so that temperature and
moisture satisfy the required saturation constraints. Because this forcing
affects pressure and density, omitting it from the preceding acoustic
substeps can excite numerical acoustic noise. Simply applying the acoustic
adjustment after microphysics would not remove the difficulty, since the
adjustment could again perturb the saturation balance. To reduce this
time-splitting error, WRF-ARW includes an estimate of the later microphysical
forcing during the dynamical integration, using information from the previous
time step \citep{skamarock2019wrf}.

This WRF-ARW architecture reflects the relevant separation of time scales:
phase transitions modify the pressure and density fields, the resulting
perturbations undergo rapid compressible adjustment, and the slower
circulation evolves under the adjusted pressure gradients. Such adjustment is
an unavoidable part of coupling phase transitions to atmospheric dynamics.
Unlike Lorenz APE, which is defined relative to a static reference state
obtained by adiabatic redistribution, the adjustment considered here takes
place within an established circulation.

It is through this adjustment that the expansion work of water vapor can be
linked to kinetic energy generation in the lower atmosphere. When water vapor
condenses and the resulting condensate precipitates, mass is removed from the
atmospheric column. In the simplest vertical setting, the associated pressure
perturbation is partly adjusted by an upward displacement of air from below,
as considered in studies of hydrostatic adjustment following rain formation
\citep[e.g.,][]{spengler11}.

In a three-dimensional circulation, this adjustment need not remain an
internal vertical redistribution within the precipitating column. If ascent
feeds an established upper-level outflow, gas-phase air displaced upward from
below can be exported horizontally from the region of precipitation. The
resulting surface-pressure deficit then contains two physically distinct
contributions: the direct loss of precipitated water mass and the additional
export of gaseous air enabled by the adjustment circulation. In this way, phase-transition work can be converted into kinetic energy
generation by horizontal pressure gradients in the lower atmosphere, with
the magnitude of this generation constrained by the available steam-engine
power.

This interpretation suggests a direct numerical test. If condensate fallout
is suppressed, so that condensed water is retained as airborne hydrometeors
rather than removed by precipitation, the column-mass sink and its associated
pressure adjustment should be weakened. If this mechanism contributes
substantially to lower-atmospheric kinetic energy generation, the resulting
circulation should weaken as well. We are not aware of analogous experiments
performed on the global scale, although disabling condensate fallout in global
climate models has recently been proposed as a diagnostic test of
condensation-sink dynamics \citep{Makarieva2025front}. For tropical cyclones,
the corresponding comparison is familiar from the reversible setting, in
which hydrometeor fallout is disabled and condensed water remains with the air
parcel. Relative to simulations with precipitation fallout, reversible
simulations produce storms that are much weaker, develop more slowly, or
exhibit substantially smaller central pressure deficits and horizontal
pressure gradients \citep[e.g.,][]{bryan09b,wang20,wang21}.

\begin{figure}
\centering

\begin{tikzpicture}[
    evidenceA/.style={
        rectangle,
        rounded corners=3pt,
        draw=black,
        thick,
        fill=yellow!20,
        align=center,
        text width=3.8cm,
        minimum height=2.6cm,
        inner sep=8pt,
        font=\small
    },
    evidenceB/.style={
        rectangle,
        rounded corners=3pt,
        draw=black,
        thick,
        fill=blue!10,
        align=center,
        text width=3.8cm,
        minimum height=2.6cm,
        inner sep=8pt,
        font=\small
    },
    block5/.style={
        rectangle,
        rounded corners=3pt,
        draw=black,
        very thick,
        fill=teal!12,
        align=center,
        minimum height=3.2cm,
        inner sep=8pt,
        font=\small\bfseries
    },
    steamBlock/.style={
        rectangle,
        rounded corners=3pt,
        draw=black,
        very thick,
        fill=red!12,
        align=center,
        minimum height=3.2cm,
        inner sep=8pt,
        font=\small\bfseries
    },
    conclusionBlock/.style={
        rectangle,
        rounded corners=3pt,
        draw=black,
        thick,
        fill=gray!15,
        align=center,
        minimum height=3.2cm,
        inner sep=8pt,
        font=\small
    },
    supportBlock/.style={
        rectangle,
        rounded corners=3pt,
        draw=black,
        thick,
        fill=green!12,
        align=center,
        minimum height=3.0cm,
        inner sep=8pt,
        font=\small
    },
    arrow/.style={
        ->,
        >=stealth,
        thick
    },
    bnum/.style={
        anchor=north west,
        font=\bfseries\small,
        inner sep=0pt
    }
]

\node[evidenceA] (m1a) {
    Mass-related\\
    pressure gradients\\
    dominate\\ in the lower atmosphere
};

\node[evidenceA, right=0.8cm of m1a] (m2a) {
    Temperature-related\\
    pressure gradients\\
    dominate\\ in the upper atmosphere
};

\node[evidenceA, right=0.8cm of m2a] (m3a) {
    Lorenz APE neglects\\
    mass-related\\
    pressure gradients\\
    by construction
};

\node[evidenceB, below=0.75cm of m1a] (m1b) {
    The lower atmosphere\\
    generates two thirds\\
    of global wind power
};

\node[evidenceB, below=0.75cm of m2a] (m2b) {
    The upper atmosphere\\
    generates one third\\
    of global wind power
};

\node[evidenceB, below=0.75cm of m3a] (m3b) {
    Differential heating\\
    generates one third\\
    of global wind power\\
    as Lorenz APE
};

\coordinate (rowthreeleft) at ([yshift=-0.95cm]m1b.south west);

\node[block5, text width=3.4cm, anchor=north west] (s5) at (rowthreeleft) {
    Steam-engine power\\
    approximates total\\
    atmospheric power
};

\node[steamBlock, text width=4.8cm, right=0.6cm of s5] (steam) {
    Wind power in the lower atmosphere\\
    arises from precipitation-driven\\
    mass-related pressure gradients
};

\node[conclusionBlock, text width=3.6cm, right=0.6cm of steam] (conc) {
    Wind power has sources beyond\\
    differential heating
};

\coordinate (rowfourleft) at ([yshift=-0.8cm]s5.south west);

\node[supportBlock, text width=4.2cm, anchor=north west] (r4left) at (rowfourleft) {
    Suppressing precipitation fallout\\
    weakens modeled storms;\\
    GCM tests are needed
};

\node[supportBlock, text width=4.0cm, right=0.55cm of r4left] (r4mid) {
    Surface pressure perturbations\\
    are of order $p_v$\\
    across weather-system scales
};

\node[supportBlock, text width=3.65cm, right=0.55cm of r4mid] (r4right) {
    Steam-engine power predicts
    maximum hurricane intensity
    as a function of $p_v$
};

\node[bnum] at ([xshift=3pt,yshift=-3pt]m1a.north west) {(a)};
\node[bnum] at ([xshift=3pt,yshift=-3pt]m2a.north west) {(b)};
\node[bnum] at ([xshift=3pt,yshift=-3pt]m3a.north west) {(c)};

\node[bnum] at ([xshift=3pt,yshift=-3pt]m1b.north west) {(d)};
\node[bnum] at ([xshift=3pt,yshift=-3pt]m2b.north west) {(e)};
\node[bnum] at ([xshift=3pt,yshift=-3pt]m3b.north west) {(f)};

\node[bnum] at ([xshift=3pt,yshift=-3pt]s5.north west) {(g)};
\node[bnum] at ([xshift=3pt,yshift=-3pt]steam.north west) {(h)};
\node[bnum] at ([xshift=3pt,yshift=-3pt]conc.north west) {(i)};

\node[bnum] at ([xshift=3pt,yshift=-3pt]r4left.north west) {(j)};
\node[bnum] at ([xshift=3pt,yshift=-3pt]r4mid.north west) {(k)};
\node[bnum] at ([xshift=3pt,yshift=-3pt]r4right.north west) {(l)};

\draw[arrow] (m1a.south) -- (m1b.north);
\draw[arrow] (m2a.south) -- (m2b.north);
\draw[arrow] (m3a.south) -- (m3b.north);

\draw[arrow] (m1b.south) -- ([xshift=-1.3cm]steam.north);

\draw[arrow] (s5.east) -- (steam.west);

\coordinate (join23) at ($(m2b.south)!0.5!(m3b.south)+(0,-0.35cm)$);
\coordinate (m2down) at ($(m2b.south)+(0,-0.35cm)$);
\coordinate (m3down) at ($(m3b.south)+(0,-0.35cm)$);

\draw[thick] (m2b.south) -- (m2down) -- (join23) -- (m3down) -- (m3b.south);
\draw[arrow] (join23) -- ([xshift=0.35cm]conc.north west);

\draw[arrow] (r4left.north) -- ([xshift=-1.5cm]steam.south);
\draw[arrow] (r4mid.north) -- (steam.south);
\draw[arrow] (r4right.north) -- ([xshift=1.5cm]steam.south);

\end{tikzpicture}

\caption{Schematic summary of evidence for a major role of steam-engine power in atmospheric energetics. In this scheme, the lower and upper atmosphere refer to the two diagnostic layers separated at $500~\mathrm{hPa}$ (Fig.~\ref{figKEvert}). ``Mass-related'' pressure gradients arise from horizontal differences in column air mass, whereas ``temperature-related'' pressure gradients arise from horizontal temperature contrasts.
(a,b) Section~\ref{sec:WK}, Table~1 of \citet{tellus17}; (c, f, i) Section~\ref{sec:lorenz}, Fig.~\ref{fig:budget}; (d, e) Section~\ref{sec:WK}, Fig.~\ref{figKEvert}; (g) Section~\ref{Wv}, Fig.~\ref{fig:budget}; (j) Section~\ref{sec:mech};
(h) Sections~\ref{sec:ciad}, \ref{sec:mech}; (k) Section~\ref{sec:ciad}; (l) \citet{mn25}.
}
\label{fig:evidence}

\end{figure}
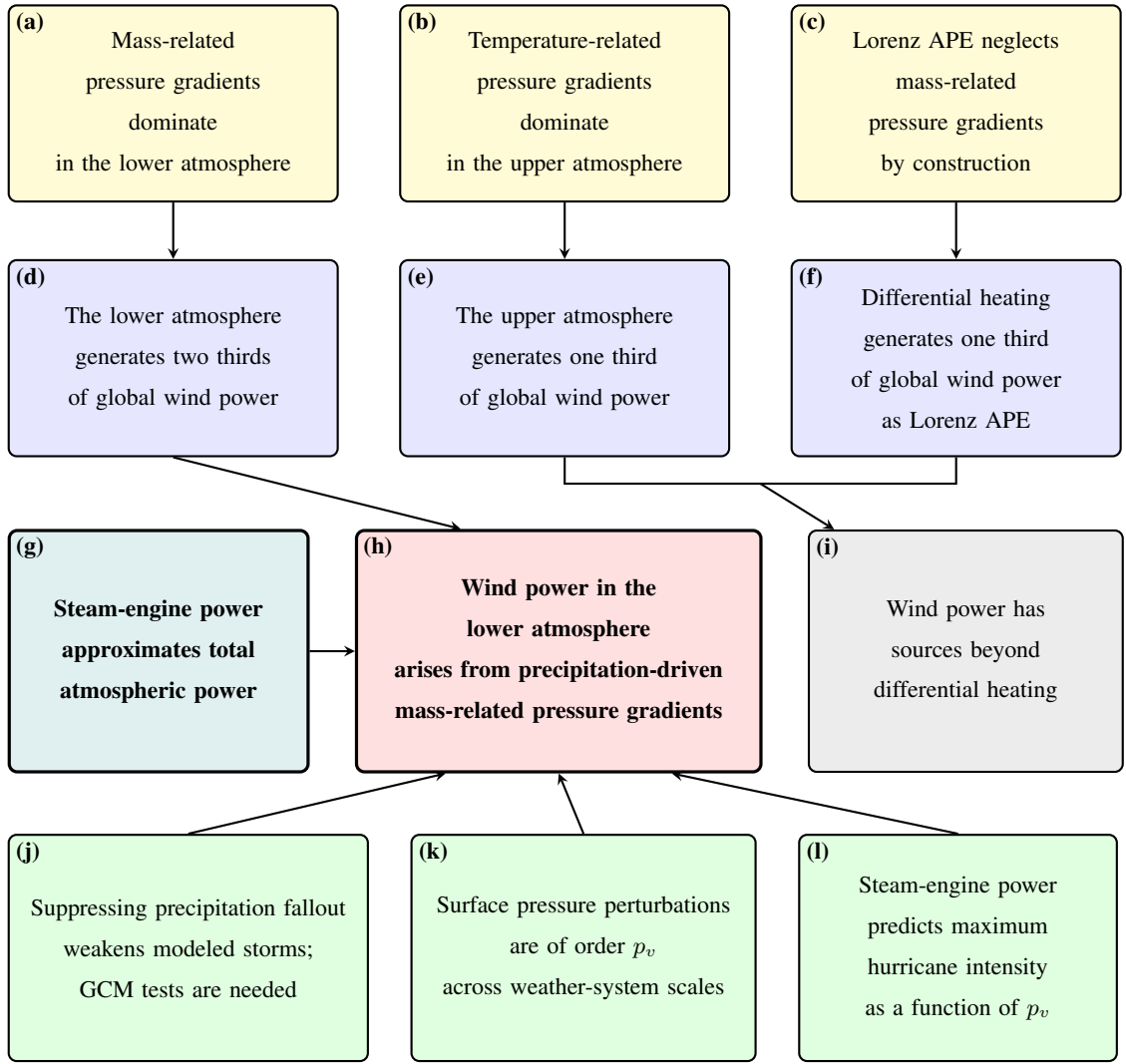

\section{Discussion}

The evidence summarized in Fig.~\ref{fig:evidence} indicates that the
steam-engine power estimated from global precipitation,
$4.4\pm0.9~\mathrm{W\,m^{-2}}$, is comparable in magnitude to the kinetic
energy generation by horizontal pressure gradients diagnosed from MERRA-2,
$W_K\simeq3.2\pm0.3~\mathrm{W\,m^{-2}}$. About two thirds of kinetic energy
is generated in the lower atmosphere, where mass-related pressure gradients
dominate, while about one third is generated aloft, where temperature-related
pressure gradients dominate. The latter contribution is comparable to Lorenz
APE generation. Thus the additional power revealed by the present analysis is
associated primarily with lower-atmospheric, mass-related pressure gradients.

The possible importance of this pathway has been obscured historically by the
static view of available potential energy. \citet{Margules1901} showed that
surface pressure perturbations contain little static available energy compared
with temperature-related pressure perturbations aloft. Lorenz APE accordingly
focused attention on differential heating and omits mass-related pressure
gradients by construction. However, small static available energy does not
imply small kinetic energy generation. In the frictional lower atmosphere,
flow crosses isobars, allowing surface pressure gradients to generate kinetic
energy directly. Aloft, by contrast, larger temperature-related pressure
gradients may yield smaller net generation because positive and negative
contributions partly cancel.

In this sense, the lower atmosphere acts as an unlocking branch of the
circulation. Temperature-related pressure gradients aloft may represent a
large reservoir of Lorenz APE, but this reservoir is not necessarily converted
into kinetic energy. In the frictionless axisymmetric problem considered by
\citet{held80}, strong upper-level pressure gradients can be maintained while
air motion remains largely along isobars, so that pressure-gradient work is
small. Surface friction changes this situation by permitting cross-isobaric
flow in the lower branch and thereby enabling the upper-level APE reservoir to
participate in kinetic energy generation.

In the terminology of \citet{bejan20}, including water vapor, phase
transitions, and precipitation fallout changes the definition of the
thermodynamic system considered. In this moist system, precipitation
introduces an additional pathway for power generation: by removing water from
the gas phase and enabling column-mass redistribution, it can create or
strengthen the surface pressure gradients driving cross-isobaric flow in the
lower atmosphere. Thus precipitation does not merely participate in unlocking
the temperature-related APE reservoir aloft; it also supplies a mass-related
source of kinetic energy generation in the lower branch itself.

The central correspondence identified here is therefore between the
condensational steam-engine contribution, $A_c\mathcal{P}$, and
lower-atmospheric kinetic energy generation, where mass-related pressure
gradients dominate. The physical significance of the difference between
$W_P$ and $W_g$ is another interesting aspect of this framework that deserves
further study.

The present analysis concerns statistically stationary circulation patterns,
in which condensation-maintained pressure gradients are expressed as kinetic
energy generation. The same framework can be extended to non-stationary
systems, where phase transitions may contribute to the formation and
intensification of pressure perturbations. In particular, the condensational power scale
$A_c\mathcal{P}\simeq RT_c\mathcal{P}$ has the same dimensions as the
vertically integrated pressure tendency in the lower inflow layer,
$\int_{\rm lower}(\partial p/\partial t)\,dz$. Whether this relation
constrains the development of storms and other transient circulations remains
to be investigated.

Finally, if condensation and precipitation contribute substantially to
atmospheric power generation, then the biospheric control of evaporation has
a direct dynamical significance. Vegetation---and especially forests---supplies
the vapor whose condensation and fallout help maintain pressure gradients and
sustain the atmospheric work that transports moisture, generates winds, and
returns water to the land as rain.

\appendix
\section{Gravitational power of precipitation}
\label{app:WP}

Here we derive Eq.~\eqref{WP}. After separating the contribution from
horizontal pressure gradients, $W_K$~\eqref{WK}, the remaining contribution to
atmospheric power is associated with vertical motion in the pressure field:
\beq\label{WPapp0}
W_P
\equiv
-\frac{1}{\mathcal{S}}
\int
w\frac{\partial p}{\partial z}\,d\Omega .
\eeq
Using hydrostatic balance, $\partial p/\partial z=-\rho g$, where $\rho$ is
the density of gaseous air, this becomes
\beq\label{WPapp1}
W_P
=
\frac{g}{\mathcal{S}}
\int \rho w\,d\Omega .
\eeq

We now relate the integral of $\rho w$ to the sources and sinks of gaseous
water. More generally, for any scalar density $\chi$ satisfying the steady
continuity equation
\[
\nabla\cdot(\chi\mathbf{v})=S_\chi ,
\]
one has
\beq\label{source_height_identity}
\int \chi w\,d\Omega
=
-\int zS_\chi\,d\Omega ,
\eeq
provided the boundary contribution from $\nabla\cdot(z\chi\mathbf{v})$ vanishes.
Indeed,
\[
\nabla\cdot(z\chi\mathbf{v})
=
z\nabla\cdot(\chi\mathbf{v})+\chi w .
\]

For the gaseous atmosphere, $\chi=\rho$ and $S_{\chi}=S$, where $S$ is the mass source
of gaseous air. Evaporation contributes positively to $S$, while condensation
removes water vapor from the gas phase. Thus Eq.~\eqref{source_height_identity}
gives
\beq\label{rhow}
\int \rho w\,d\Omega
=
-\int zS\,d\Omega .
\eeq

Let $E$ and $C$ be the local mass rates of evaporation and condensation, in
$\mathrm{kg\,m^{-3}\,s^{-1}}$, both defined positive. Then
\beq\label{Ssource}
S=E-C .
\eeq
Substituting Eq.~\eqref{Ssource} into Eq.~\eqref{rhow} gives
\beq\label{rhowEC}
-\frac{1}{\mathcal{S}}\int zS\,d\Omega
=
\frac{1}{\mathcal{S}}
\left(
\int zC\,d\Omega
-
\int zE\,d\Omega
\right).
\eeq

Define the mass precipitation flux $P$ by
\beq\label{Papp}
P
\equiv
\frac{1}{\mathcal{S}}\int C\,d\Omega
=
\frac{1}{\mathcal{S}}\int E\,d\Omega ,
\eeq
where the equality holds in steady state. The mean condensation and evaporation
heights are then
\beq\label{zcapp}
z_c
\equiv
\frac{1}{\mathcal{S}P}
\int zC\,d\Omega,
\quad
z_1
\equiv
\frac{1}{\mathcal{S}P}
\int zE\,d\Omega .
\eeq
Therefore
\beq\label{rhowzc}
-\frac{1}{\mathcal{S}}\int zS\,d\Omega
=
P(z_c-z_1).
\eeq
Combining Eqs.~\eqref{WPapp1}, \eqref{rhow}, and \eqref{rhowzc}, we obtain  Eq.~\eqref{WP}.

\section{Diagnosing $W_K$ from MERRA-2}
\label{app:WK}

We diagnose kinetic energy generation by horizontal pressure gradients from
three-hourly MERRA-2 pressure-level data. The calculation uses the horizontal
wind components $u$ and $v$, the height $Z$ of each isobaric surface, surface
pressure $p_s$, and the pressure-level coordinate $p$. The geopotential of an
isobaric surface is
\begin{equation}
\label{Z}
\Phi(p,\varphi,\lambda,t) = g Z(p,\varphi,\lambda,t),
\end{equation}
where $g=9.80665~\mathrm{m\,s^{-2}}$. In pressure coordinates, the horizontal
pressure-gradient force per unit mass is $-\nabla_p \Phi$. Thus the local rate
of kinetic energy generation per unit mass is
\begin{equation}
c_K(p,\varphi,\lambda,t)
=
-\mathbf{u}\cdot\nabla_p\Phi
=
-\left(
u\frac{\partial \Phi}{\partial x}
+
v\frac{\partial \Phi}{\partial y}
\right),
\label{eq:ck_local}
\end{equation}
with units $\mathrm{W\,kg^{-1}}$. This is the pressure-coordinate form of the
local work term $-\mathbf{u}\cdot\nabla p$ in Eq.~\eqref{WK}.

The horizontal derivatives in Eq.~\eqref{eq:ck_local} were evaluated by
centered finite differences on each pressure surface. For longitude, periodic
boundary conditions were used:
\begin{equation}
\frac{\partial \Phi}{\partial x}
\simeq
\frac{1}{R_E\cos\varphi}
\frac{\Phi(p,\varphi,\lambda+\Delta\lambda)
      -\Phi(p,\varphi,\lambda-\Delta\lambda)}
     {2\Delta\lambda}.
\label{eq:dphidx}
\end{equation}
For latitude,
\begin{equation}
\frac{\partial \Phi}{\partial y}
\simeq
\frac{\Phi(p,\varphi+\Delta\varphi,\lambda)
      -\Phi(p,\varphi-\Delta\varphi,\lambda)}
     {2R_E\Delta\varphi}.
\label{eq:dphidy}
\end{equation}
Here $R_E=6371~\mathrm{km}$ is Earth's radius, and angular increments are in
radians. Near the poles, where $\cos\varphi$ is numerically small, the zonal
derivative was not used.

Values were treated as invalid if they were non-numeric or exceeded $10^{14}$
in absolute value, which excludes the MERRA-2 missing-value flag near
$10^{15}$. A value of $c_K$ was computed only where both wind components and
the neighboring geopotential values required for the finite differences were
valid. Otherwise the grid point was marked as undefined.

At each horizontal grid point, the resolved pressure-column integral was
computed by trapezoidal integration over adjacent valid pressure levels:
\begin{equation}
K_0(\varphi,\lambda,t)
=
\sum_{k=1}^{N_p-1}
\frac{1}{2}
\left(c_{K,k}+c_{K,k+1}\right)
\frac{\Delta p_k}{g},
\label{eq:column_base}
\end{equation}
where $\Delta p_k=|p_{k+1}-p_k|$. Since $\Delta p/g$ is the atmospheric mass
per unit area between two pressure surfaces, $K_0$ has units
$\mathrm{W\,m^{-2}}$. Figure~\ref{figKEvert} shows the vertical distribution
of this pressure-layer contribution for 2023.

Because the lowest valid pressure level does not always coincide with the
local surface pressure, we computed two surface-layer corrections. Let $p_b$
be the lowest valid pressure level in a column, $c_b$ the corresponding value
of $c_K$, and $\Delta p_s=p_s-p_b$. If $\Delta p_s>0$, the correction used for
the main estimate assumes that $\mathbf{u}\cdot\nabla p=0$ at the surface and
that the mean local kinetic energy generation between $p_b$ and $p_s$ is
one half of its value at $p_b$:
\begin{equation}
K_{\mathrm{corr}}
=
K_0
+
\frac{1}{2}c_b\frac{\Delta p_s}{g}.
\label{eq:surface_corr}
\end{equation}
This is the value reported in the main text.

As a sensitivity estimate, we also extrapolated $c_K$ to the surface from the
two lowest valid pressure levels. If $p_u$ is the next pressure level above
$p_b$, with value $c_u$, then
\begin{equation}
c_s
=
c_b
+
(c_b-c_u)
\frac{p_s-p_b}{p_b-p_u},
\label{eq:surface_extrap_ck}
\end{equation}
and
\begin{equation}
K_{\mathrm{extrap}}
=
K_0
+
\frac{1}{2}
(c_b+c_s)
\frac{\Delta p_s}{g}.
\label{eq:surface_extrap}
\end{equation}
If the second valid level was unavailable, the extrapolated correction was
replaced by Eq.~\eqref{eq:surface_corr}.

Finally, global means were obtained by area-weighting the column values:
\begin{equation}
\overline{K}(t)
=
\frac{\sum_{i,j} K_{ij}(t) A_{ij}}
     {\sum_{i,j} A_{ij}},
\label{eq:global_mean}
\end{equation}
where the grid-cell area was computed from latitude-band boundaries,
\begin{equation}
A_{ij}
=
R_E^2\Delta\lambda
\left[
\sin\left(\varphi_{i+1/2}\right)
-
\sin\left(\varphi_{i-1/2}\right)
\right].
\end{equation}
The output contains three global time series: the resolved pressure-level
integral $\overline{K}_0$, the half-bottom-value surface correction
$\overline{K}_{\mathrm{corr}}$, and the extrapolated surface correction
$\overline{K}_{\mathrm{extrap}}$. The value discussed in the paper is
\[
W_K = \overline{K}_{\mathrm{corr}} .
\]
The difference between $\overline{K}_{\mathrm{corr}}$ and
$\overline{K}_{\mathrm{extrap}}$ is used to characterize the sensitivity of
$W_K$ to the unresolved surface layer.

\section{Derivation of Eq.~\eqref{WvapP}}
\label{app:Wvap}

Here we derive Eq.~\eqref{WvapP} using the same per-unit-area convention as in
Eq.~\eqref{W}. With boundary fluxes neglected, Eq.~\eqref{Wvapgrad} gives
\beq\label{WvapDt}
W_{\rm vap}
=
-\frac{1}{\mathcal{S}}
\int
\frac{D p_v}{D t}\,d\Omega ,
\eeq
where $D/Dt\equiv \partial/\partial t+\mathbf{v}\cdot\nabla$ and steady state
has been assumed.

Let $\mathcal{N}_d$ be the dry-air molar density and
$d\mathcal{M}_d\equiv \mathcal{N}_d\,d\Omega$ the dry-air molar element. Using
$p_d=\mathcal{N}_dRT$ and $\gamma\equiv p_v/p_d$, Eq.~\eqref{WvapDt} becomes
\beq\label{WvapM}
W_{\rm vap}
=
-\frac{1}{\mathcal{S}}
\int
\frac{\gamma RT}{p_v}
\frac{D p_v}{D t}\,
d\mathcal{M}_d .
\eeq
This is the power analogue of Eq.~\eqref{INT}: $dp_v^*/p_v^*$ is replaced by
$(Dp_v/Dt)/p_v$, and the integral is taken over the whole atmosphere rather
than only over the condensation layer.

Using $p_v=\gamma p_d$, Eq.~\eqref{WvapM} separates into
\beq\label{Wvapsplit}
W_{\rm vap}
=
-\frac{1}{\mathcal{S}}
\int
\frac{\gamma RT}{p_d}
\frac{D p_d}{D t}\,
d\mathcal{M}_d
-
\frac{1}{\mathcal{S}}
\int
RT
\frac{D\gamma}{D t}\,
d\mathcal{M}_d .
\eeq

We first evaluate the pressure term. Since
$d\mathcal{M}_d=\mathcal{N}_d\,d\Omega$ and $p_d=\mathcal{N}_dRT$,
\beq\label{Wvappress1}
-\frac{1}{\mathcal{S}}
\int
\frac{\gamma RT}{p_d}
\frac{D p_d}{D t}\,
d\mathcal{M}_d
=
-\frac{1}{\mathcal{S}}
\int
\gamma
\frac{D p_d}{D t}\,
d\Omega .
\eeq
Following the approximation used by \citet[][p.~149]{pauluis2002b}, we retain
the vertical pressure-gradient contribution,
\beq\label{vertapprox}
\frac{D p_d}{D t}
=
\mathbf{v}\cdot\nabla p_d
\simeq
w\frac{\partial p}{\partial z}.
\eeq
With hydrostatic balance, $\partial p/\partial z=-\rho g\simeq
-M\mathcal{N}_d g$, where $M$ is the molar mass of dry air, this gives
\beq\label{Wvappress2}
-\frac{1}{\mathcal{S}}
\int
\gamma
\frac{D p_d}{D t}\,
d\Omega
\simeq
\frac{Mg}{\mathcal{S}}
\int
\mathcal{N}_v w\,d\Omega ,
\eeq
where $\mathcal{N}_v=\gamma\mathcal{N}_d$.

Applying the source--height identity Eq.~\eqref{source_height_identity} to
water vapor, with $\chi=\mathcal{N}_v$ and
$S_\chi=\dot{\mathcal{N}}_e-\dot{\mathcal{N}}_c$, gives
\beq\label{Nvwid}
\int
\mathcal{N}_v w\,d\Omega
=
\int
z\dot{\mathcal{N}}_c\,d\Omega
-
\int
z\dot{\mathcal{N}}_e\,d\Omega ,
\eeq
where $\dot{\mathcal{N}}_e$ and $\dot{\mathcal{N}}_c$ are the molar rates of
evaporation and condensation per unit volume, both defined positive. If
evaporation occurs at $z_1=0$, the evaporation term vanishes. Defining the
molar precipitation flux $\mathcal{P}$ and the mean condensation height $z_c$
by
\beq\label{PzcWvap}
\mathcal{P}
=
\frac{1}{\mathcal{S}}\int \dot{\mathcal{N}}_c\,d\Omega,
\quad
z_c
=
\frac{1}{\mathcal{S}\mathcal{P}}
\int z\dot{\mathcal{N}}_c\,d\Omega ,
\eeq
we obtain
\beq\label{Wvappress3}
-\frac{1}{\mathcal{S}}
\int
\frac{\gamma RT}{p_d}
\frac{D p_d}{D t}\,
d\mathcal{M}_d
\simeq
Mgz_c\mathcal{P}.
\eeq
For $z_1=0$, this is identical to the lifting contribution
$A_g\mathcal{P}$ in Eq.~\eqref{ft}.

We now evaluate the second term in Eq.~\eqref{Wvapsplit}. From
$\gamma=\mathcal{N}_v/\mathcal{N}_d$, dry-air continuity, and vapor
continuity,
\beq\label{DgammaEC}
\frac{D\gamma}{D t}
=
\frac{\dot{\mathcal{N}}_e-\dot{\mathcal{N}}_c}{\mathcal{N}_d}.
\eeq
Therefore,
\beq\label{Wvapsec1}
-\frac{1}{\mathcal{S}}
\int
RT
\frac{D\gamma}{D t}\,
d\mathcal{M}_d
=
-\frac{1}{\mathcal{S}}
\int
RT(\dot{\mathcal{N}}_e-\dot{\mathcal{N}}_c)\,d\Omega .
\eeq
Introducing the evaporation- and condensation-weighted mean temperatures
\beq\label{TbTcWvap}
T_b
\equiv
\frac{1}{\mathcal{S}\mathcal{P}}
\int T\dot{\mathcal{N}}_e\,d\Omega,
\quad
T_c
\equiv
\frac{1}{\mathcal{S}\mathcal{P}}
\int T\dot{\mathcal{N}}_c\,d\Omega ,
\eeq
we find
\beq\label{Wvapsec2}
-\frac{1}{\mathcal{S}}
\int
RT(\dot{\mathcal{N}}_e-\dot{\mathcal{N}}_c)\,d\Omega
=
R(T_c-T_b)\mathcal{P}.
\eeq
Unlike the condensational contribution $A_c\mathcal{P}$ in
Eq.~\eqref{Av2}, this term includes the evaporation region and therefore
contains the negative contribution $-RT_b\mathcal{P}$.

Combining Eqs.~\eqref{Wvapsplit}, \eqref{Wvappress3}, and
\eqref{Wvapsec2}, we obtain Eq.~\eqref{WvapP}.

\dataavailability{The MERRA-2 instantaneous three-hourly pressure-level
meteorological fields used in this study
(\texttt{inst3\_3d\_asm\_Np}; M2I3NPASM, version 5.12.4; 1980--2025)
are available from NASA GES DISC at
\url{https://doi.org/10.5067/QBZ6MG944HW0}. The GPCP Precipitation Level 3
Monthly 0.5-Degree data (version 3.3; \texttt{GPCPMON}; 1983--2025) are
available from NASA GES DISC at
\url{https://doi.org/10.5067/MEASURES/GPCP/DATA306}.}

\codeavailability{The code used to diagnose kinetic energy generation from
MERRA-2 and to prepare the corresponding derived results will be deposited in
Zenodo. The Zenodo DOI will be added in a subsequent version of this preprint.}


\end{document}